\documentclass[
  ,draft            
  ]
  {aipproc}

\layoutstyle{6x9}

\usepackage{amssymb,amsmath}

\newcommand{\dback}{D_{\rm b}}
\newcommand{\dbacktilde}{{\tilde D}_{\rm b}}

\newcommand{\DD}{{\mathcal D}}

\newcommand{\figcite}[1]{{\protect \cite{#1}}}

\newcommand{\normfig}{0.7\textwidth}

\newcommand{\cref}[1]{Chapter~\ref{#1}}

\newcommand{\fref}[1]{Fig.~\ref{#1}}

\newcommand{\phd}{{\protect \vphantom{\dagger}}}
\newcommand{\hc}{{\rm h.c.} }
\newcommand{\sign}[1]{{\rm Sign}(#1)}
\newcommand{\vf}{v_F}

\newcommand{\step}{Y}

\newcommand{\kf}{k_F}

\begin{document}

\title{Strong correlations in low dimensional systems}

\author{T. Giamarchi}{
  address={University of Geneva, 24 Quai Ernest Ansermet, 1211 Geneva, Switzerland}
}

\begin{abstract}
I describe in these notes the physical properties of one dimensional
interacting quantum particles. In one dimension the combined effects
of interactions and quantum fluctuations lead to a radically new
physics quite different from the one existing in the higher
dimensional world. Although the general physics and concepts are
presented, I focuss in these notes on the properties of interacting
bosons, with a special emphasis on cold atomic physics in optical
lattices. The method of bosonization used to tackle such problems is
presented. It is then used to solve two fundamental problems. The
first one is the action of a periodic potential, leading to a
superfluid to (Mott)-Insulator transition. The second is the action
of a random potential that transforms the superfluid in phase
localized by disorder, the Bose glass. Some discussion of other
interesting extensions of these studies is given.
\end{abstract}

\maketitle


\section{Introduction and choice of contents}

One-dimensional systems of interacting particles are particularly
fascinating both from a theoretical and experimental point of view.
Such systems have been extensively investigated theoretically for
more than 40 years now. They are wonderful systems in which
interactions play a very special role and whose physics is
drastically different from the `normal' physics of interacting
particles, that is, the one known in higher dimensions. From the
theoretical point of view they present quite unique features. The
one-dimensional character makes the problem simple enough so that
some rather complete solutions could be obtained using specific
methods, and yet complex enough to lead to incredibly rich physics
\cite{emery_revue_1d,solyom_revue_1d,voit_bosonization_revue,schulz_houches_revue,%
vondelft_bosonization_review,schonhammer_bosonization_review,senechal_bosonization_revue,%
gogolin_1dbook,giamarchi_book_1d}.

Crucial theoretical progress were made and many theoretical tools
got developed, mostly in the 1970's allowing a detailed
understanding of the properties of such systems. This culminated in
the 1980's with a new concept of interacting one-dimensional
particles, analogous to the Fermi liquid for interacting electrons
in three dimensions: the Luttinger liquid. Since then many
developments have enriched further our understanding of such
systems, ranging from conformal field theory to important progress
in the exact solutions such as Bethe ansatz
\cite{takahashi_book_bethe}.

In addition to these important theoretical progress, experimental
realizations have knows comparably spectacular developments.
One-dimensional systems were mostly at the beginning a theorist's
toy. Experimental realizations started to appear in the 1970's with
polymers and organic compounds. But in the last 20 years or so we
have seen a real explosion of realization of one-dimensional
systems. The progress in material research made it possible to
realize bulk materials with one-dimensional structures inside. The
most famous ones are the organic superconductors
\cite{review_organics_complete} and the spin and ladder compounds
\cite{dagotto_ladder_review}. At the same time, the tremendous
progress in nanotechnology allowed to obtain realizations of
isolated one-dimensional systems such as quantum wires
\cite{fisher_transport_luttinger_review}, Josephson junction arrays
\cite{fazio_josephson_junction_review}, edge states in quantum hall
systems \cite{wen_edge_review}, and nanotubes
\cite{dresselhaus_book_fullerenes_nanotubes}. Last but not least,
the recent progress in Bose condensation in optical traps have
allowed an unprecedented way to probe for strong interaction effects
in such systems \cite{pitaevskii_becbook,greiner_mott_bec}.

The goal of these lectures was therefore to present the major
theoretical tools of the domain. However, while writing these notes,
I was faced with a dilemma. Having written a recent book on this
very subject \cite{giamarchi_book_1d} it felt that writing these
notes would be a simple repetition or worse a butchering of the
explanations that could be found in the book. I have thus chosen to
give to these notes a slightly different focuss than the material
that was actually presented during the course. For an introduction
to the one dimensional systems and in particular for the fermionic
problems, as well as most of the technical details, I refer the
reader to \cite{giamarchi_book_1d} where all this material is
described in detail and hopefully in a pedagogical fashion suitable
for graduate students. I have chosen to restrict these notes to the
description of one and quasi-one dimensional systems of bosons.
Indeed the spectacular recent progress made thanks to cold atomic
gases, make it useful to have a short summary on the subject, in
complement of the material that can already be found in
\cite{giamarchi_book_1d}. Note that although these notes do cover
some of the basic material for cold atoms they cannot pretend to be
an exhaustive and complete review on this rapidly developing
subject. The whole volume of this book would not be sufficient for
that. Rather they reflect a partial selection, based on my own
excitement in the field and its connections with the low dimensional
world. I thus apologize in advance for those whose pet theory,
experiment or paper I would fail to mention in these notes.

\section{One dimensional bosons, and their peculiarities} \label{sec:1dbosons}

Bosons are particularly interesting systems to investigate. From the
theoretical point of view bosons present quite interesting
peculiarities and are in fact a priori much more difficult to treat
than their fermionic counterpart. Indeed, for fermions, the free
fermion approximation is usually a good starting point, at least in
high enough dimension where Fermi liquid theory holds. Some
perturbations such as disorder can be studied for the much simpler
free fermion case, the Pauli principle ensuring that even in the
absence of interactions the perturbation remains small compared to
the characteristic scales of the free problem (here the Fermi
energy). One can thus gain valuable physical intuition on the
problem before adding the interactions. For bosons, on the contrary,
interactions are needed from the start since there are radical
differences between a non-interacting boson gas and an interacting
one. Bosons have another remarkable property, namely in the absence
of interactions all the particles can condense in a macroscopic
state. Interacting bosons thus constitute a remarkable theoretical
challenge. One dimension presents additional peculiarities as we
will see below.

Before embarking on the subject of interacting bosons, let us first
discuss how one can obtain ``one dimensional'' objects. Of course
the real world is three dimensional, but all the one dimensional
system are characterized by a confining potential forcing the
particles to be in a localized states
\begin{figure}
  \centerline{\includegraphics[width=\normfig]{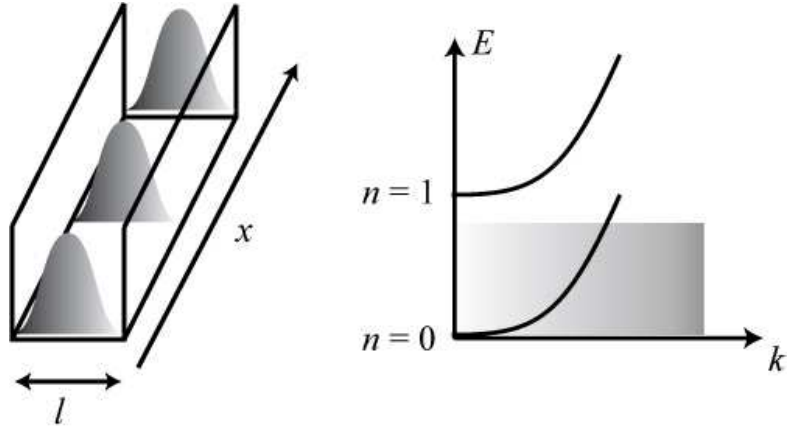}}
 \caption{(left) Confinement of the electron gas in a one-dimensional tube of transverse size $l$.
 $x$ is the direction of the tube. Only one transverse direction of
 confinement has been shown for clarity. Due to the transverse confining
 potential the transverse degrees of freedom are strongly quantized.
 (right) Dispersion relation $E(k)$. Only half of the dispersion relation is shown for clarity.
 $k$ is the momentum parallel to the tube direction.
 The degrees of freedom transverse to the tube direction lead to the formation of minibands,
 labelled by a quantum number $n$. One can be in a situation where only one miniband can be excited,
 due to temperature or interactions, the energy scale of which is
 represented by the gray box. In that case the system is equivalent
 to a one dimensional system where only longitudinal degrees of
 freedom can vary.}
 \label{fig:confwire}
\end{figure}
The wavefunction of the system is thus of the form
\begin{equation}
 \psi(x,y) = e^{i k x} \phi(y)
\end{equation}
where $\phi$ depends on the precise form of the confining potential
For an infinite well, a show in \fref{fig:confwire}, $\phi$ is
$\phi(y) = \sin((2n_y+1)\pi y/l)$, whereas it would be a gaussian
function (\ref{eq:harmfond}) for an harmonic confinement. The energy
is of the form
\begin{equation}
 E = \frac{k^2}{2 m} + \frac{k_y^2}{2 m}
\end{equation}
where for simplicity I have taken hard walls confinement. The
important point is the fact that due to the narrowness of the
transverse channel $l$, the quantization of $k_y$ is sizeable.
Indeed, the change in energy by changing the transverse quantum
number $n_y$ is at least (e.g. $n_y =0$ to $n_y=1$)
\begin{equation}
 \Delta E = \frac{3\pi^2}{2 m l^2}
\end{equation}
This leads to minibands as shown in \fref{fig:confwire}. If the
distance between the minibands is larger than the temperature or
interactions energy one is in a situation where only one miniband
can be excited. The transverse degrees of freedom are thus frozen
and only $k$ matters. The system is a one-dimensional quantum
system.

We can thus forget about the transverse directions and model the
bosons keeping only the longitudinal degrees of freedom. Two
slightly different starting points are possible. One can start
directly in the continuum, where bosons are described by
\begin{equation} \label{eq:boscont}
 H = \int dx \frac{\hbar^2(\nabla\psi)^\dagger(\nabla\psi)}{2m} +
 \frac12\int dx \;dx'\; V(x-x') \rho(x)\rho(x') - \mu_0 \int dx\; \rho(x)
\end{equation}
The first term is the kinetic energy, the second term is the
repulsion $V$ between the bosons and the last term is the chemical
potential. A famous model, exactly solvable by Bethe-ansatz
\cite{lieb_bosons_1D} uses a local repulsion
\begin{equation} \label{eq:localint}
 V(x) = V_0 \delta(x)
\end{equation}
Note that $V$ is not the real atom-atom interaction in three
dimensional space but an effective interaction where the transverse
degrees of freedom have already been incorporated. The extension of
the transverse wavefunction is of course much larger than the
dimensions of the atoms themselves, so the hard core repulsion
between two atoms can be safely forgotten. The three dimensional
interaction is characterized by a scattering length
\cite{pitaevskii_becbook} $a_s$ by
\begin{equation}
 V(x,y,z) = \frac{4\pi\hbar^2a_s}{m}\delta(x)\delta(y)\delta(z)
\end{equation}
Note that for bosons the s-wave scattering is the important one
since two bosons can get close together due to the symmetry of their
wave function, while for fermions this scattering would be highly
inefficient. $V_0$ in (\ref{eq:localint}) is obtained by integrating
over the transverse degrees of freedom of the wavefunction. Because
the effective interaction depends on the extension of the transverse
wavefunction it will be possible to vary (increase) it by increasing
the confinement \cite{olshanii_potential_bec}. To be closer to the
situation for cold atomic gases one has to remember that there is
also a confining potential in the longitudinal direction even if
this one is much more shallow and therefore the chemical potential
is spatially dependent, leading to a term of the form
\begin{equation} \label{eq:varchem}
 H = \int dx [V_c(x) - \mu_0] \rho(x)
\end{equation}
where $\mu(x) = \frac12 m \omega_0^2 x^2$ is the confining
potential. In the absence of interactions the ground state of the
system is given by an harmonic oscillator wavefunction
\begin{equation} \label{eq:harmfond}
 \psi_0(x) = \left(\frac{m \omega_0}{\hbar \pi }\right)^{1/4} e^{-\frac{m \omega_0}{2 \hbar} x^2}
\end{equation}
In the absence of the confining potential the bosons are in a plane
wave state of momentum $k=0$, whereas here they are confined on a
typical length of order $a_K = \sqrt{\hbar/(m\omega_0)}$ in presence
of the harmonic potential. A similar but much tighter confinement is
imposed in the transverse directions as well, leading to the
formation of the tubes as discussed above. Typical longitudinal
lengths due to the harmonic confinement are $15 \mu m$ while the
transverse dimensions can be $60 nm$
\cite{stoferle_tonks_optical,kinoshita_tonks_continuous,paredes_tonks_optical}.

Due to the trap the density profile is thus non homogeneous. In the
absence of interactions it would just be the gaussian profile of
(\ref{eq:harmfond}). In presence of interactions a similar effect
occurs but the profile changes. A very simple way to see this effect
is when one can neglect the kinetic energy (so called Thomas-Fermi
approximation; for other situations see \cite{pitaevskii_becbook}).
In that case the density profile is obtained by minimizing
(\ref{eq:boscont}) and (\ref{eq:varchem}), leading to
\begin{equation}
 V_0 \rho(x) + [V_c(x)-\mu_0] = 0
\end{equation}
the density profile is thus an inverted parabola, reflecting the
change of the chemical potential. One can express the confining
length as $a = a_K \sqrt{(2 \rho_0 V_0)/(\hbar \omega_0)}$ where
$\rho_0$ is the density at the center of the trap. Of course dealing
with such inhomogeneous system is a complication and has
consequences that I will discuss below. To treat this problem, there
are various approximations that one can make. The crudest way of
dealing with such a confinement is simply to ignore the spatial
variation and remember the trap as giving a finite size to the
system.

In the model (\ref{eq:boscont}), the bosons move in a continuum. It
is interesting to add to the system (\ref{eq:boscont}) a periodic
potential $V_L(x)$ coupled to the density
\cite{jaksch_bose_hubbard,greiner_mott_bec,stoferle_tonks_optical}
\begin{equation} \label{eq:boslat}
 H_L = \int dx\; V_L(x) \rho(x)
\end{equation}
This term, which favors certain points in space for the position of
the bosons, mimics the presence of a lattice of period $a$, the
periodicity of the potential $V_L(x)$. We take the potential as
\begin{equation}
 V_L(x) = V_L \sin^2(k x) = \frac{V_L}2 [1-\cos(2 k x)]
\end{equation}
one has thus $a = \pi/k$. The presence of the lattice can
drastically change the properties of an interacting one dimensional
system as I will discuss below.

If the lattice is much higher than the kinetic energy it is better
to start from a tight binding representation
\cite{ziman_solid_book}. In that case in each minima of the lattice
one can approximate the periodic potential by an harmonic one
$\frac12 (4V_L) k^2 x^2$. One has thus on each site harmonic
oscillator wavefunctions that hybridize to form a band. If $V_L$ is
large the energy levels in each well are well separated and one can
retain only the ground state wavefunction in each well. The system
can then be represented directly by a model defined on a lattice
\begin{equation} \label{eq:bosehub}
 H = -t\sum_i (b^\dagger_{i+1} b^\phd_i+ \hc) + U\sum_i n_i(n_i-1)
  - \sum_i \mu_i n_i
\end{equation}
where $b^\phd_i$ (resp $b^\dagger_i$) destroys (resp. creates) a
boson on site $i$. The parameters $t$, $U$, and $\mu_i$ are
respectively the effective hopping, interaction and local chemical
potential. Because the overlap between different sites is very small
the interaction is really local. Since atoms are neutral this model
is a very good approximation of the experimental situation. Such a
model known as a Bose-Hubbard model has been used extensively in a
variety of other contexts (see e.g. \cite{giamarchi_book_1d} for
more details and references). The effective parameters $t$ and $U$
can be easily computed  by a standard tight binding calculation
using the shape of the on site wave function (\ref{eq:harmfond})
with $\omega_0^2 = 4 V_L k^2$
\begin{equation}
\begin{split}
 t & = \langle \psi_0(x+a) | H_{\rm kin} | \psi_0(x) \rangle \\
 U & = \int dxdydz |\psi(x,y,z)|^4
\end{split}
\end{equation}
where $\psi(x,y,z) = \psi_0(x)\psi_\perp(y)\psi_\perp(z)$ and
$\psi_\perp$ is identical to (\ref{eq:harmfond}) but with $V_L$
replaced by the transverse confinement $V_\perp$. For large lattice
sizes an approximate formula is given by
\cite{zwerger_JU_expressions}
\begin{equation} \label{eq:opticalparam}
 \begin{split}
  J/E_r &= (4/\sqrt{\pi}) (V_L/E_r)^{(3/4)} \exp{(-2
  \sqrt{V_L/E_r})} \\
  U/E_r &= 4 \sqrt{2\pi}(a_s/2a) (V_L V_\perp^2/E_r^3)^{(1/4)}
 \end{split}
\end{equation}
Here $E_r = \hbar^2 k^2/(2m)$ is the so called recoil energy, i.e.
the kinetic energy for a momentum of order $\pi/a$. $V_\perp$ the
denotes the harmonic confining potential in the two transverse
directions of the tube. Typical values for the above parameters are
$a_s \sim 5 nm$ while $a \sim 400 nm$ \cite{stoferle_tonks_optical}.
The repulsion term acts if there are two or more bosons per site. It
is easy to see from (\ref{eq:opticalparam}) that, in addition to the
special effects created by the lattice itself, imposing an optical
lattice is a simple way to kill the kinetic energy of the system
while leaving interactions practically unaffected. It is thus a
convenient way to make the quantum system ``more interacting'' and
has been used as such. Of course, it is possible to also add to
(\ref{eq:bosehub}) longer range interactions if they are present in
the microscopic system. One naively expects the two models
(\ref{eq:boscont}) plus the lattice terms (\ref{eq:boslat}) and
(\ref{eq:bosehub}) to have the same asymptotic physics, the latter
one being of course much more well suited in the case of large
periodic potential.

Of course the above models are very difficult to solve, since the
tools that one usually uses fail because of the one dimensional
nature of the problem. It is customary when dealing with a
superfluid to use a Ginzburg-Landau (GL) mean field theory where the
order parameter $\Psi(x)$ represents the condensed fraction. The
time dependent GL is the celebrated Gross-Pitaevskii (GP) equation
\cite{pitaevskii_becbook}. However in one dimension it is impossible
to break a continuum symmetry even at zero temperature so a true
condensate cannot exist for an infinite size system \footnote{In
presence of the trap a condensate can exist, simply because of the
finite size effect \cite{petrov_BEC_finitesize}.}. This means that
quantum fluctuation will play an important role and that the GP
equation is not a very good starting point. One has thus to find
other ways to deal with the interactions. The model
(\ref{eq:boscont}) in the continuum is exactly solvable by Bethe
ansatz (BA) \cite{lieb_bosons_1D}, which provides very useful
physical insight. Unfortunately the BA solution does not allow the
calculation of quantities such as asymptotic correlation functions,
and thus must be supplemented by other techniques. For the
particular model of bosons with a local repulsion, one point of
special interest is the point where the repulsion between the bosons
is infinite. The system is then known as hard core bosons. In that
case it becomes impossible to put two bosons on the same site. This
is the Tonks-Girardeau (TG) limit
\cite{girardeau_tonks_gas,lieb_bosons_1D}. It is easy to see that in
that case this system of hard core bosons can be mapped either to a
spin chain system (the presence or absence of bosons being
respectively an up or down spin), or by a Jordan-Wigner
transformation to a system of spinless fermions. We will use
repeatedly this analogy between hard core bosons and fermions and
the following sections. More details on the various mappings and
equivalences between spins, fermions and bosons can be found in
\cite{giamarchi_book_1d}.

\section{Bosonization technique} \label{sec:bozo}

Treating interacting bosons in one dimension is a quite difficult
task. One very interesting technique is provided by the so-called
bosonization. It has the advantage of giving a very simple
description of the low energy properties of the system, and of being
completely general and very useful for many one dimensional systems.
This chapter will thus describe it in some details. For more details
and physical insights on this technique both for fermions and bosons
I refer the reader to \cite{giamarchi_book_1d}.

\subsection{Bosonization dictionary}

The idea behind the bosonization technique is to reexpress the
excitations of the system in a basis of collective excitations
\cite{haldane_bosons}. Indeed in one dimension it is easy to realize
that single particle excitations cannot really exit. One particle
when moving will push its neighbors and so on, which means that any
individual motion is converted into a collective one. One can thus
hope that a base of collective excitations is a good basis to
represent the excitations of a one dimensional system.

\begin{figure}
  \centerline{\includegraphics[width=\normfig]{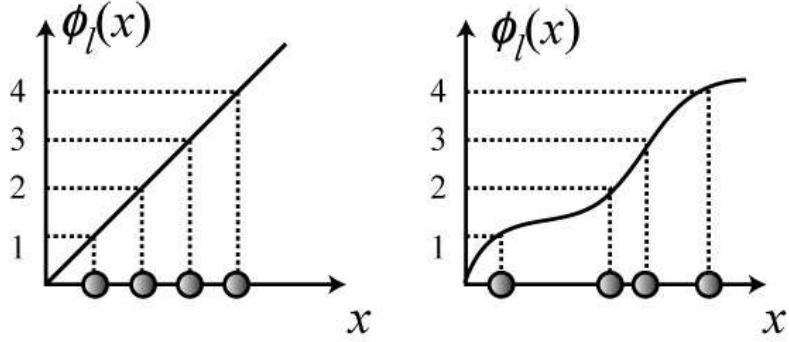}}
 \caption{Some examples of the labelling field $\phi_l(x)$. If the
 particles form a perfect lattice of lattice spacing $d$, then
 $\phi_l^0(x) = 2\pi x/d$, and is just a straight line.
 Different functions
 $\phi_l(x)$ allow to put the particles at any position in space. Note that $\phi(x)$ is always
 an increasing function regardless of the position of the particles.}
 \label{fig:labelfield}
\end{figure}

To exploit this idea, let us start with the density operator
\begin{equation}\label{eq:densmoche}
 \rho(x) = \sum_i \delta(x-x_i)
\end{equation}
where $x_i$ is the position operator of the $i$th particle. We label
the position of the $i$th particle by an `equilibrium' position
$R_i^0$ that the particle would occupy if the particles were forming
a perfect crystalline lattice, and the displacement $u_i$ relative
to this equilibrium position. Thus,
\begin{equation}
 x_i = R_i^0 + u_i
\end{equation}
If $\rho_0$ is the average density of particles, $d=\rho_0^{-1}$ is
the distance between the particles. Then, the equilibrium position
of the $i$th particle is
\begin{equation}
 R_i^0 = d i
\end{equation}
Note that at that stage it is not important whether we are dealing
with fermions or bosons. The density operator written as
(\ref{eq:densmoche}) is not very convenient. To rewrite it in a more
pleasant form we introduce a labelling field $\phi_l(x)$
\cite{haldane_bosons}. This field, which is a continuous function of
the position, takes the value $\phi_l(x_i) = 2\pi i$ at the position
of the $i$th particle. It can thus be viewed as a way to number the
particles. Since in one dimension, contrary to higher dimensions,
one can always number the particles in an unique way (e.g. starting
at $x=-\infty$ and processing from left to right), this field is
always well-defined. Some examples are shown in
\fref{fig:labelfield}.
Using this labelling field and the rules for transforming $\delta$
functions
\begin{equation}
 \delta(f(x)) = \sum_{\mbox{zeros of $f$}} \frac1{|f'(x_i)|}
 \delta(x-x_i)
\end{equation}
one can rewrite the density as
\begin{eqnarray}
 \rho(x) &=& \sum_i \delta(x-x_i) \nonumber \\
  &=& \sum_n |\nabla \phi_l(x)| \delta(\phi_l(x) - 2\pi n)
  \label{eq:denslab}
\end{eqnarray}
It is easy to see from \fref{fig:labelfield} that $\phi_l(x)$ can
always be taken as an increasing function of $x$, which allows to
drop the absolute value in (\ref{eq:denslab}). Using the Poisson
summation formula this can be rewritten
\begin{equation}
 \rho(x) = \frac{\nabla \phi_l(x)}{2\pi} \sum_p e^{i p \phi_l(x)}
\end{equation}
where $p$ is an integer. It is convenient to define a field $\phi$
relative to the perfect crystalline solution and to introduce
\begin{equation}
 \phi_l(x) = 2 \pi \rho_0 x - 2\phi(x)
\end{equation}
The density becomes
\begin{equation} \label{eq:locintbos}
 \rho(x) = \left[\rho_0 -\frac{1}{\pi} \nabla \phi(x)\right] \sum_p e^{i 2 p (\pi \rho_0 x - \phi(x))}
\end{equation}
Since the density operators at two different sites commute it is
normal to expect that the field $\phi(x)$ commutes with itself. Note
that if one averages the density over distances large compared to
the interparticle distance $d$ all oscillating terms in
(\ref{eq:locintbos}) vanish. Thus, only $p=0$ remains and this
smeared density is
\begin{equation} \label{eq:smeardens}
 \rho_{q\sim 0}(x) \simeq \rho_0 - \frac1\pi \nabla\phi(x)
\end{equation}
We can now write the single-particle creation operator
$\psi^\dagger(x)$. Such an operator can always be written as
\begin{equation} \label{eq:singlephen}
 \psi^\dagger(x) = [\rho(x)]^{1/2}e^{-i \theta(x)}
\end{equation}
where $\theta(x)$ is some operator. In the case where one would have
Bose condensation, $\theta$ would just be the superfluid phase of
the system. The commutation relations between the $\psi$ impose some
commutation relations between the density operators and the
$\theta(x)$. For bosons, the condition is
\begin{equation} \label{eq:comphen}
 [\psi^\phd_B(x),\psi_B^\dagger(x')] = \delta(x-x')
\end{equation}
Using (\ref{eq:singlephen}) the commutator gives
\begin{equation} \label{eq:comexpl}
 e^{+i \theta(x)}[\rho(x)]^{1/2}[\rho(x')]^{1/2}e^{-i \theta(x')}
 -[\rho(x')]^{1/2}e^{-i \theta(x')}e^{+i \theta(x)}[\rho(x)]^{1/2}
\end{equation}
If we assume quite reasonably that the field $\theta$ commutes with
itself ($[\theta(x),\theta(x')] = 0$), the commutator
(\ref{eq:comexpl}) is obviously zero for $x \ne x'$ if (for $x\ne
x'$)
\begin{equation} \label{eq:comcompl}
 [[\rho(x)]^{1/2},e^{-i \theta(x')}] = 0
\end{equation}
A sufficient condition to satisfy (\ref{eq:comphen}) would thus be
\begin{equation}\label{eq:commutbas}
 [\rho(x),e^{-i\theta(x')}] = \delta(x-x')e^{-i\theta(x')}
\end{equation}
It is easy to check that if the density were only the smeared
density (\ref{eq:smeardens}) then (\ref{eq:commutbas}) is obviously
satisfied if
\begin{equation} \label{eq:conjphi}
 [\frac1\pi \nabla\phi(x),\theta(x')] = -i \delta(x-x')
\end{equation}
One can show that this is indeed the correct condition to use
\cite{giamarchi_book_1d}. Equation (\ref{eq:conjphi}) proves that
$\theta$ and $\frac1\pi \nabla\phi$ are canonically conjugate. Note
that for the moment this results from totally general considerations
and does not rest on a given microscopic model. Such commutation
relations are also physically very reasonable since they encode the
well known duality relation between the superfluid phase and the
total number of particles. Integrating by part (\ref{eq:conjphi})
shows that
\begin{equation}
 \pi\Pi(x) = \hbar\nabla\theta(x)
\end{equation}
where $\Pi(x)$ is the canonically conjugate momentum to $\phi(x)$.

To obtain the single-particle operator one can substitute
(\ref{eq:locintbos}) into (\ref{eq:singlephen}). Since the square
root of a delta function is also a delta function up to a
normalization factor the square root of $\rho$ is identical to
$\rho$ up to a normalization factor that depends on the ultraviolet
structure of the theory. Thus,
\begin{equation} \label{eq:singlebos}
 \psi^\dagger_B(x) = [\rho_0 - \frac1\pi\nabla \phi(x)]^{1/2}
 \sum_{p} e^{i 2 p (\pi \rho_0 x - \phi(x))}e^{-i \theta(x)}
\end{equation}
where the index $B$ emphasizes that this is the representation of a
\emph{bosonic} creation operator. A similar formula can be derived
for fermionic operators \cite{giamarchi_book_1d}. The above formulas
are a way to represent the excitations of the system directly in
terms of variables defined in the continuum limit, and
(\ref{eq:singlebos}) and (\ref{eq:locintbos}) are the basis of the
bosonization dictionary.

The fact that all operators are now expressed in terms of variables
describing \emph{collective} excitations is at the heart of the use
of such representation, since as already pointed out, in one
dimension excitations are necessarily collective as soon as
interactions are present. In addition the fields $\phi$ and $\theta$
have a very simple physical interpretation. If one forgets their
canonical commutation relations, order in $\theta$ indicates that
the system has a coherent phase as indicated by
(\ref{eq:singlebos}), which is the signature of superfluidity. On
the other hand order in $\phi$ means that the density is a perfectly
periodic pattern as can be seen from (\ref{eq:locintbos}). This
means that the system of bosons has ``crystallized''. As we now see,
the simplicity of this representation in fact allows to solve an
interacting system of bosons in one dimension.

\subsection{Physical results and Luttinger liquid}

What is the Hamiltonian of the system? Using (\ref{eq:singlebos}),
the kinetic energy becomes
\begin{equation}
 H_K \simeq \int dx \frac{\hbar^2\rho_0}{2m}(\nabla e^{i\theta})(\nabla e^{-i\theta})
     = \int dx \frac{\hbar^2\rho_0}{2m} (\nabla\theta)^2
\end{equation}
which is the part coming from the single-particle operator
containing less powers of $\nabla\phi$ and thus the most relevant.
Using (\ref{eq:boscont}) and (\ref{eq:locintbos}), the interaction
term becomes
\begin{equation} \label{eq:intbosbos}
 H_{\rm int} = \int dx V_0 \frac{1}{2\pi^2} (\nabla\phi)^2
\end{equation}
plus higher order operators. Keeping only the above lowest order
shows that the Hamiltonian of the interacting bosonic system can be
rewritten as
\begin{equation} \label{eq:luthamphen}
 H = \frac{\hbar}{2\pi}\int dx [\frac{u K}{\hbar^2} (\pi \Pi(x))^2 + \frac{u}{K}
 (\nabla\phi(x))^2]
\end{equation}
where I have put back the $\hbar$ for completeness. This leads to
the action
\begin{equation} \label{eq:lutacphen}
 S/\hbar = \frac1{2\pi K} \int dx \;d\tau [\frac1u (\partial_\tau\phi)^2 +
 u (\partial_x\phi(x))^2]
\end{equation}
This hamiltonian is a standard sound wave one. The fluctuation of
the phase $\phi$ represent the ``phonon'' modes of the density wave
as given by (\ref{eq:locintbos}). One immediately sees that this
action leads to a dispersion relation, $\omega^2 = u^2 k^2$, i.e. to
a linear spectrum. $u$ is the velocity of the excitations. $K$ is a
dimensionless parameter whose role will be apparent below.  The
parameters $u$ and $K$ are used to parameterize the two coefficients
in front of the two operators. In the above expressions they are
given by
\begin{equation} \label{eq:ukpert}
\begin{split}
 u K & = \frac{\pi  \hbar \rho_0}m \\
 \frac{u}{K} &= \frac{V_0}{\hbar \pi }
\end{split}
\end{equation}
This shows that for weak interactions $u \propto (\rho_0 V_0)^{1/2}$
while $K \propto (\rho_0 / V_0)^{1/2}$. In establishing the above
expressions we have thrown away the higher order operators, that are
less relevant. The important point is that these higher order terms
will not change the form of the Hamiltonian (like making cross terms
between $\phi$ and $\theta$ appears etc.) but {\it only} renormalize
the coefficients $u$ and $K$ (for more details see
\cite{giamarchi_book_1d}). For galilean invariant system the first
relation is exactly satisfied regardless of the strength of the
interaction
\cite{haldane_bosons,cazalilla_correlations_1d,giamarchi_book_1d}.

The low-energy properties of interacting bosons are thus described
by an Hamiltonian of the form (\ref{eq:luthamphen}) \emph{provided}
the proper $u$ and $K$ are used. These two coefficients
\emph{totally} characterize the low-energy properties of massless
one-dimensional systems. The bosonic representation and Hamiltonian
(\ref{eq:luthamphen}) play the same role for one-dimensional systems
than the Fermi liquid theory plays for higher-dimensional systems.
It is an effective low-energy theory that is the fixed point of any
massless phase, regardless of the precise form of the microscopic
Hamiltonian. This theory, which is known as Luttinger liquid theory
\cite{haldane_bosonisation,haldane_bosons}, depends only on the two
parameters $u$ and $K$. Provided that the correct value of these
parameters are used, \emph{all} asymptotic properties of the
correlation functions of the system then can be obtained
\emph{exactly} using (\ref{eq:locintbos}) and (\ref{eq:singlephen}).

In the absence of a good perturbation theory (e.g. in the
interaction) such as (\ref{eq:ukpert}), it is difficult to compute
these coefficients. One has two ways of proceeding. Either one is
attached to a particular microscopic model (such as the Bose-Hubbard
model for example). In which case the Luttinger liquid coefficients
$u$ and $K$ are functions of the microscopic parameters. One thus
just needs two relations involving these coefficients that can be
computed with the microscopic model and determine these
coefficients, thus allowing to compute {\emph all} correlation
functions. How to do that depends on taste and integrability or not
of the model. If the model is integrable by Bethe-ansatz such as the
Lieb-Liniger model one computes thermodynamics from BA and obtains
$u$ and $K$ that way
\cite{haldane_bosons,cazalilla_correlations_1d}. If the model is not
exactly solvable one can still use numerics such as exact
diagonalization, monte-carlo or DMRG technique to compute these
coefficients. Because they can be extracted from thermodynamic
quantities, their determination suffers usually from very little
finite size effects compared to a direct calculation of the
correlation functions. The Luttinger liquid theory thus provides,
coupled with the numerics, an incredibly accurate way to compute
correlations and physical properties of a system. For more details
on the various procedures and models see \cite{giamarchi_book_1d}.

But, of course, a much more important use of Luttinger liquid theory
is to justify the use of the boson Hamiltonian and fermion--boson
relations as starting points for any microscopic model. The
Luttinger parameters then become some effective parameters. They can
be taken as input, based on general rules (e.g. for bosons
$K=\infty$ for non interacting bosons and $K$ decreases as the
repulsion increases, for other general rules see
\cite{giamarchi_book_1d}), without any reference to a particular
microscopic model. This removes part of the caricatural aspects of
any modelization of a true experimental system. This use of the
Luttinger liquid is reminiscent of the one made of Fermi liquid
theory. Very often calculations are performed in solids starting
from `free' electrons and adding important perturbations (such as
the BCS attractive interaction to obtain superconductivity). The
justification of such a procedure is rooted in the Fermi liquid
theory, where one does not deal with `real' electrons but with the
quasiparticles, which are intrinsically fermionic in nature. The
mass $m$ and the Fermi velocity $\vf$ are then some parameters. The
calculations in $d=1$ proceed in the same spirit with the Luttinger
liquid replacing the Fermi liquid. The Luttinger liquid theory is
thus an invaluable tool to tackle the effect of perturbations on an
interacting one-dimensional electron gas (such as the effect of
lattice, impurities, coupling between chains, etc.). I will
illustrate such use in the following sections, taking as examples
the effects of a periodic potential and a disordered one.

Let us now examine in details the physical properties of such a
Luttinger liquid. For this we need the correlation functions. I
briefly show here how to compute them using the standard operator
technique. More detailed calculations and functional integral
methods are given in \cite{giamarchi_book_1d}. A building block to
compute the various observables is
\begin{equation} \label{eq:correlgphi}
 G_{\phi\phi}(x,\tau) = \langle T_\tau [\phi(x,\tau)-\phi(0,0)]^2 \rangle
\end{equation}
where $T_\tau$ is the standard time ordering operator, and $\tau$
the imaginary time \cite{mahan_book}. We absorb the factor $K$ in
the Hamiltonian by rescaling the fields (this preserves the
commutation relation)
\begin{equation}
\begin{split}
 \phi &= \sqrt{K} \tilde\phi \\
 \theta &= \frac{1}{\sqrt{K}} \tilde\theta
\end{split}
\end{equation}
The fields $\tilde\phi$ and $\tilde\theta$ can be expressed in terms
of bosons operator $[b^\phd_q,b^\dagger_{q'}]=\delta_{q,q'}$. This
ensures that their canonical commutation relations are satisfied.
One has
\begin{equation} \label{eq:phitetbos}
\begin{split}
 \phi(x) &=  - \frac{i\pi}L
 \sum_{p\ne 0} \left(\frac{L |p|}{2\pi}\right)^{1/2}
 \frac1p e^{-\alpha|p|/2-i p x}(b^\dagger_p + b^\phd_{-p}) \\
 \theta(x) &=  \frac{i\pi}L
 \sum_{p\ne 0} \left(\frac{L |p|}{2\pi}\right)^{1/2}
 \frac1{|p|} e^{-\alpha|p|/2-i p x}(b^\dagger_p - b^\phd_{-p})
\end{split}
\end{equation}
where $L$ is the size of the system and $\alpha$ a short distance
cutoff (of the order of the interparticle distance) needed to
regularize the theory at short scales. The above expressions are in
fact slightly simplified and zero modes should also be incorporated
\cite{giamarchi_book_1d}. This will not affect the remaining of this
section and the calculation of the correlation functions.

It is easy to check by a direct substitution of (\ref{eq:phitetbos})
in (\ref{eq:luthamphen}) that Hamiltonian (\ref{eq:luthamphen}) with
$K=1$ is simply
\begin{equation} \label{eq:hamsimlut}
 \tilde H = \sum_{p \ne 0} u |p| b^\dagger_p b^\phd_p
\end{equation}
The time dependence of the field can now be easily computed from
(\ref{eq:hamsimlut}) and (\ref{eq:phitetbos}). This gives
\begin{equation} \label{eq:timdepphi}
 \phi(x,\tau) =  - \frac{i\pi}L
 \sum_{p\ne 0} \left(\frac{L |p|}{2\pi}\right)^{1/2}
 \frac1p e^{-\alpha|p|/2-i p x}(b^\dagger_p e^{u|p|\tau} + b^\phd_{-p} e^{-u|p|\tau})
\end{equation}
The correlation function (\ref{eq:correlgphi}) thus becomes
\begin{align}
 G_{\phi\phi}(x,\tau) &= K \langle T_\tau [\tilde\phi(x,\tau)-\tilde\phi(0,0)]^2 \rangle_0 \nonumber\\
 &= 2 K[\langle \tilde\phi(0,0)\tilde\phi(0,0) \rangle_0 -
 \step(\tau)\langle \tilde\phi(x,\tau)\tilde\phi(0,0) \rangle_0 \nonumber \\
 & \quad - \step(-\tau)\langle \tilde\phi(0,0)\tilde\phi(x,\tau) \rangle_0]
 \label{eq:correlgphimod}
\end{align}
where $\step$ is the step function. One then plugs
(\ref{eq:timdepphi}) in (\ref{eq:correlgphimod}). The calculation is
thus reduced to the averages of factors such as
\begin{equation}
 \langle b^\dagger_p b^\phd_{p'} \rangle_0 = \delta_{p,p'} f_B(\epsilon_p = u|p|)
\end{equation}
and factors such as $b b^\dagger = 1 - b^\dagger b$ that can be
easily reduced to the above form. $f_B$ is the standard Bose factor.
At $T=0$ since $\epsilon_p >0$ (remember that $p \ne 0$ for the
bosons modes) $f_B(\epsilon_q) = 0$. Thus, (\ref{eq:correlgphimod})
becomes (taking the standard limit $L\to \infty$)
\begin{eqnarray}
 G_{\phi\phi}(x,\tau) &=& K \int_0^\infty \frac{dp}p e^{-\alpha p}
 [1 - e^{-u |\tau| p} \cos(p x)]
 \nonumber \\
 &=& \frac{K}{2} \log\left[\frac{x^2 + (u |\tau|+\alpha)^2}{\alpha^2}\right]
\end{eqnarray}
Thus, up to the small cutoff $\alpha$, this is essentially $\log(r)$
where $r$ is the distance in space--time. This invariance by
rotation in space--time reflects the Lorentz invariance of the
action. One can introduce
\begin{equation}
\begin{split}
 r &= \sqrt{x^2 + y_\alpha^2} \\
 y_\alpha &= u \tau + \alpha\sign{\tau}
\end{split}
\end{equation}
The same calculation with $\theta$ instead of $\phi$ gives exactly
the same result with $1/K$ instead of $K$. One can either do it
directly or notice that the Hamiltonian is invariant by $\phi\to
\theta$ and $K \to 1/K$. The above calculations have been performed
at zero temperature. It is easy to obtain the correlation at finite
temperature using the same methods. It can also be derived using the
conformal invariance of the theory. Such conformal invariance can
also be nicely used to obtain the correlations for systems of finite
size
\cite{cazalilla_finitesize_luttinger,cazalilla_correlations_1d}.
Other correlations and further details can be found in
\cite{giamarchi_book_1d}.

\begin{figure}
 \centerline{\includegraphics[width=\normfig]{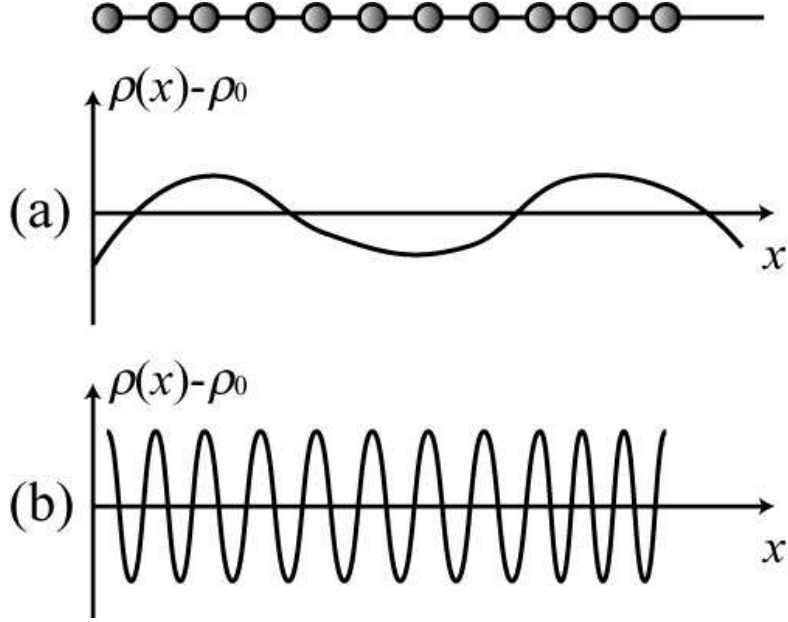}}
 \caption{The density $\rho(x)$ can be decomposed in components varying with different Fourier
 wavevectors. The characteristic scale to separate these modes is the inter particle
 distance. Only the two lowest harmonics are represented here.
 Although they have very different spatial variations both these
 modes depends on the \emph{same} smooth field $\phi(x)$.
 (a) the smooth variations of the density at lengthscale larger than the lattice spacing. These are simply
 $-\nabla\phi(x)/\pi$. (b) The density wave corresponding to oscillations of the density at a wavevector $Q=2\pi\rho_0$. These modes
 correspond to the operator $e^{i\pm2\phi(x)}$.}
 \label{fig:densmodes}
\end{figure}

In order to compute physical observable we need to get correlations
of exponentials of the fields $\phi$ and $\theta$. To do so one
simply uses that for an operator $A$ that is \emph{linear} in terms
of boson fields and a quadratic Hamiltonian one has
\begin{equation}
 \langle T_\tau e^{A} \rangle = e^{\frac{1}{2} \langle T_\tau A^2 \rangle}
\end{equation}
Thus, for example
\begin{eqnarray}
 \langle T_\tau e^{i2\phi(x,\tau)}e^{-i2\phi(0,0)}\rangle &=&
 e^{-2\langle T_\tau [\phi(x,\tau)-\phi(0,0)]^2\rangle}
 \nonumber \\
 &=& e^{-2 G_{\phi\phi}(x,\tau)}
\end{eqnarray}
If we want to compute the fluctuations of the density
\begin{equation}
\langle T_\tau \rho(x,\tau) \rho(0) \rangle
\end{equation}
we obtain, using (\ref{eq:locintbos})
\begin{multline}\label{eq:densluttinger}
 \langle T_\tau \rho(x,\tau) \rho(0) \rangle = \rho_0^2
 + \frac{K}{2\pi^2}
 \frac{y_\alpha^2-x^2}{(x^2+y_\alpha^2)^2}
 + \rho_0^2 A_2 \cos(2\pi\rho_0 x) \left(\frac{\alpha}{r}\right)^{2K}
 \\
 + \rho_0^2 A_4 \;\cos(4\pi\rho_0 x) \left(\frac{\alpha}{r}\right)^{8K} + \cdots
\end{multline}
Here, the lowest distance in the theory is $\alpha \sim
\rho_0^{-1}$. The amplitudes $A_i$ are non-universal objects. They
depend on the precise microscopic model, and even on the parameters
of the model. Contrary to the amplitudes $A_n$, which depend on the
precise microscopic model, the power-law decay of the various terms
are \emph{universal}. They \emph{all} depend on the unique Luttinger
coefficient $K$. Physically the interpretation of the above formula
is that the density of particles has fluctuations that can be sorted
compared to the average distance between particles $\alpha \sim d =
\rho_0^{-1}$. This is shown on \fref{fig:densmodes}.
The fluctuations of long wavelength decay with a universal power
law. These fluctuations correspond to the hydrodynamic modes of the
interacting boson fluid. The fact that their fluctuation decay very
slowly is the signature that there are massless modes present. This
corresponds to the sound waves of density described by
(\ref{eq:luthamphen}). However the density of particles has also
higher fourier harmonics. The corresponding fluctuations also decay
very slowly but this time with a non-universal exponent that is
controlled by the LL parameter $K$. This is also the signature of
the presence of a continuum of gapless modes, that exists for
Fourier components around $Q = 2n\pi \rho_0$ as shown in
\fref{fig:densmodes}. In the Tonks-Girardeau limit, this mode is
simply the low energy mode corresponding to transferring one fermion
from one side of the Fermi surface to the other, leading to a $2\kf$
momentum transfer. In higher dimensions and with a true condensate
such a gapless mode would not exist, and only the modes close to
$q\sim 0$ would remain (the Goldstone modes corresponding to the
phase fluctuations). The other gapless mode is thus the equivalent
of the roton minimum that only exists at a finite energy in high
dimensions but would be pushed to zero energy in a one dimensional
situation
\cite{nozieres_roton,iucci_absorption,cazalilla_deconfinement_long}.
As we discussed the coefficient $K$ goes to infinity when the
interaction goes to zero which means that the correlations in the
density decays increasingly faster with smaller interactions. This
is consistent with the idea that the system becoming more and more
superfluid smears more and more its density fluctuations.

\begin{figure}
  \centerline{\includegraphics[width=\normfig]{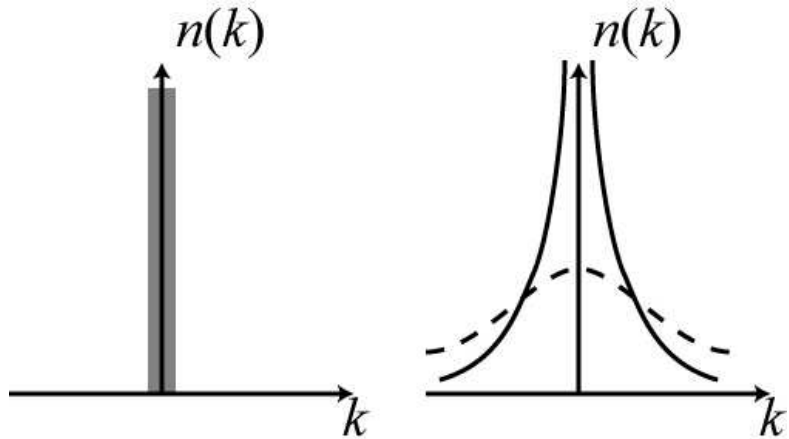}}
 \caption{Momentum distribution $n(k)$ for the bosons as a function of the momentum $k$.
 (left) For non-interacting bosons all bosons are in $k=0$ state, thus $n(k)\propto \delta(k)$ (thick line).
 (right) As soon as interactions are introduced a true condensate
 cannot exist. The $\delta$ function is replaced by a power law
 divergence with an exponent $\nu = 1-1/(2K)$ (solid line). In a Mott insulating phase or a Bose glass phase,
 the superfluid correlation functions decay exponentially leading to a rounding of the divergence and a lorentzian like
 shape for $n(k)$. This is indicated by the dashed line.}
 \label{fig:singlebos}
\end{figure}

Let us now turn to the single particle correlation function
\begin{equation}
 G(x,\tau) = \langle T_\tau \psi(x,\tau) \psi^\dagger(0,0) \rangle
\end{equation}
At equal time this correlation function is a direct measure on
whether a true condensate exists in the system. Its Fourier
transform is the occupation factor $n(k)$. In presence of a true
condensate, this correlation function tends to $G(x\to \infty,\tau
=0) \to |\psi_0|^2$ the square of the order parameter $\psi_0 =
\langle \psi(x,\tau) \rangle$ when there is superfluidity. Its
Fourier transform is a delta function at $q=0$, as shown in
\fref{fig:singlebos}. In one dimension, no condensate can exist
since it is impossible to break a continuous symmetry even at zero
temperature, so this correlation must always go to zero for large
space or time separation. Using (\ref{eq:singlebos}) the correlation
function can easily be computed. Keeping only the most relevant term
($p=0$) leads to (I have also put back the density result for
comparison)
\begin{equation} \label{eq:singleboscor}
\begin{split}
  \langle T_\tau \psi(r) \psi^\dagger(0)\rangle &= A_1
  \left(\frac{\alpha}{r}\right)^{\frac1{2K}} + \cdots \\
  \langle T_\tau \rho(r)\rho(0)\rangle &= \rho_0^2 + \frac{K}{2\pi^2}
  \frac{y_\alpha^2 - x^2}{(y_\alpha^2 + x^2)^2} +
  A_3 \cos(2\pi\rho_0 x) \left(\frac{1}{r}\right)^{2K} + \cdots
\end{split}
\end{equation}
where the $A_i$ are the non-universal amplitudes. For the
non-interacting system $K=\infty$ and we recover that the system
possesses off-diagonal long-range order since the single-particle
Green's function does not decay with distance. The system has
condensed in the $q=0$ state. As the repulsion increases ($K$
decreases), the correlation function decays faster and the system
has less and less tendency towards superconductivity. The occupation
factor $n(k)$ has thus no delta function divergence but a power law
one, as shown in \fref{fig:singlebos}.
Note that the presence of the condensate or not is not directly
linked to the question of superfluidity. The fact that the system is
a Luttinger liquid with a finite velocity $u$, implies that in one
dimension an interacting boson system has always a linear spectrum
$\omega= u k$, contrary to a free boson system where $\omega \propto
k^2$. Such a system is thus a \emph{true} superfluid at $T=0$ since
superfluidity is the consequence of the linear spectrum
\cite{mikeska_supra_1d}. Note that of course when the interaction
tends to zero $u\to 0$ as it should to give back the quadratic
dispersion of free bosons.

\begin{figure}
  \centerline{\includegraphics[width=\normfig]{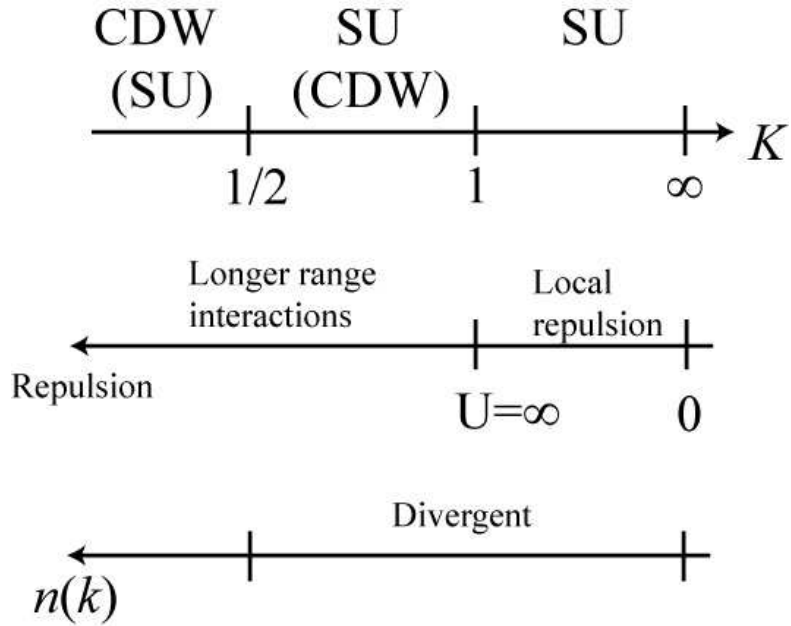}}
 \caption{``Phase diagram'' for (incommensurate) one-dimensional bosons, as a function
 of the parameter $K$. The phase indicated corresponds to the most
 divergent susceptibility, while a phase in parenthesis corresponds
 to a subdominant divergence. The value $K=1$ corresponds to the Tonks-Girardeau limit
 where the bosons are hard core and behave very similarly to spinless fermions.
 The region where $n(k)$ has a singularity at $k=0$ is indicated on the bottom graph.}
 \label{fig:phasebos}
\end{figure}

An even better criterion for the occurrence of superfluidity or
other ordered phases is provided by the susceptibilities. They are
the Fourier transforms of the correlation functions
\begin{equation}
 \chi(\omega_n,k) = \int dx d\tau \chi(x,\tau)
\end{equation}
It is easy to see by simple dimensional analysis that if the
correlation decays as a powerlaw $\chi(r) \sim (1/r)^\mu$ then the
susceptibility behaves as $\max(\omega,k,T)^{\mu-2}$. The
susceptibilities give direct indications on the phase that the
system would tend to realize, if many chains were put together and
coupled by a mean-field interaction. An RPA calculation would then
directly lead to the stabilization of three dimensional order,
stabilizing the phase with the most divergent susceptibility. From
(\ref{eq:singleboscor}) the charge and superfluid susceptibilities
diverge as
\begin{equation}
\begin{split}
 \chi_\rho &= T^{2-2K} \\
 \chi_\psi &= T^{2-\frac1{2K}}
\end{split}
\end{equation}
This leads thus to the ``phase diagram'' of \fref{fig:phasebos}.
Let me again emphasize that there is no true long range order in the
system but only algebraically decaying correlations. Such a phase
diagram indicates the dominant tendency of the system. Note also
that the superfluid susceptibility is \emph{not} identical to
$n(k)$, since this one only contains the correlation at equal time.
Its divergence is different and, as shown in \fref{fig:singlebos} is
given by
\begin{equation}
 n(k) \propto k^{\frac1{2K}-1}
\end{equation}
As we already discussed, for a purely local interaction, when the
repulsion becomes infinite the system becomes equivalent to free
spinless fermions. Indeed two particles cannot be on the same site
and the particles are totally free except for this constraint. In
that case the decay of the density (\ref{eq:singleboscor}) should be
the one of free fermions, i.e. $1/r^2$. This can be realized if
$K=1$. Note that the Green function of the bosons \emph{does not}
become the correlation function of spinless fermions since they
still represent different statistics. In particular the boson
correlation function still diverges at $k=0$ even in the TG limit.
In that limit since $K=1$ the $n(k)$ has a square root divergence.
For a purely local repulsion, $K=1$ is the minimal value that $K$
can reach. Of course, longer range repulsion between bosons can make
the system reach smaller values of $K$. More details and mapping on
other systems (classical and quantum such as spin chains) can be
found in \cite{giamarchi_book_1d}. Testing these predictions in cold
atomic systems is complicated by the presence of the harmonic trap
\cite{stoferle_tonks_optical,kinoshita_tonks_continuous,paredes_tonks_optical,%
cazalilla_tonks_gases}

\section{Mott transition} \label{sec:mott}

\subsection{Basic Ideas}

\begin{figure}
  \centerline{\includegraphics[width=\normfig]{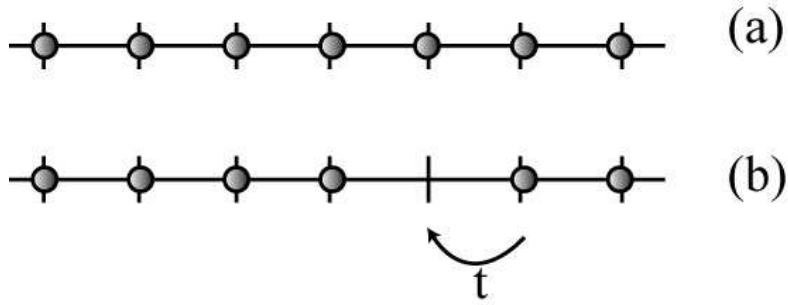}}
 \caption{(a) If there is one particle per lattice site and the repulsion among particles
 is strong, a plane wave state for the particles is energetically
 unfavorable since the density is uniform. It is better to localize the particles on each site.
 Such a state is a Mott insulator since hopping would cost an energy of the order of the interaction
 $U$ among particles. (b) If the system is doped the extra particles or holes can propagate without
 any energy cost from the interactions and gain some kinetic energy \ $\sim\!t$. The system is then
 in general superfluid again.}
 \label{fig:mottcart}
\end{figure}

\begin{figure}
  \centerline{\includegraphics[width=\normfig]{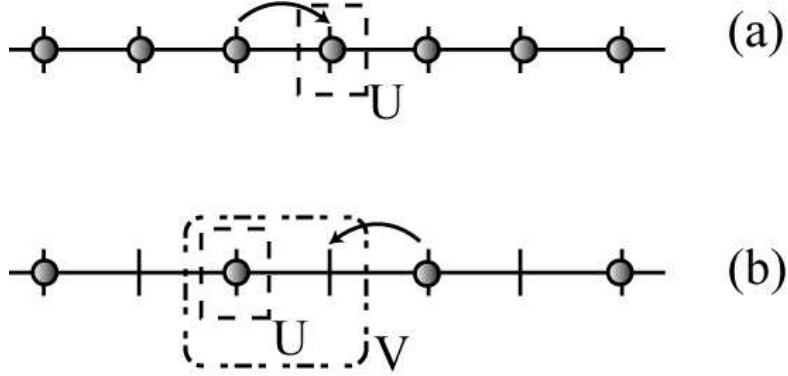}}
 \caption{(a) In a model with onsite interactions ($U$) a Mott
 insulator only exists for one particle per site. (b) Nearest neighbor
 interactions $V$ can stabilize the insulating state for up to one
 particle every two sites. And so on with longer ranger interactions.
 The longer the range of the interaction, the higher the commensurability for which one
 can have a Mott insulator provided the interactions are large enough.}
 \label{fig:quartmot}
\end{figure}

Let us now investigate the effects of a lattice on such a bosonic
system
\cite{haldane_bosons,fisher_boson_loc,scalettar_bosons,%
giamarchi_attract_1d,giamarchi_mott_shortrev,buchler_cic_bec}. For
noninteracting bosons the lattice just provides a renormalization of
the kinetic energy as shown in (\ref{eq:opticalparam}). When
interactions are present, a lattice leads to a radically new
physics. In particular when the density of carriers is commensurate
with the lattice, another interaction induced phenomenon occurs. In
that case the system can become an insulator. This is the mechanism
known as Mott transition
\cite{mott_historical_insulator,mott_metal_insulator}, and is a
metal-insulator transition induced by the interactions. The physics
of a Mott insulator is well-known and illustrated in
\fref{fig:mottcart}.
If the repulsion $U$ among the particles is much larger than the
kinetic energy $t$, then the plane wave state is not very favorable
since it leads to a uniform density where particles experience the
maximum repulsion. It is more favorable to localize the particles on
the lattice sites to minimize the repulsion and the system is an
insulator for one particle per site. If the system is weakly doped
compared to a state with one particle per site the holes can
propagate without experiencing repulsion, the system is thus in
general a superfluid again but with a number of carriers
proportional to the doping. The above argument shows that, in high
dimensions, one usually needs a finite (and in general of the order
of the kinetic energy) repulsion to reach that state. For further
details on the Mott transition in higher dimension see
\cite{georges_d=infini,imada_mott_review}. It is important to note
that one particle per site is not the only commensurate filling
where one can in principle get a Mott insulator, but that every
commensurate filling can work, in principle, depending on the
interactions. This is illustrated in \fref{fig:quartmot}.
It is indeed easy to see that for large enough onsite ($U$) and
nearest neighbor ($V$) repulsion a quarter-filled system is an
ordered Mott insulator. As I will discuss in more details below, and
as is clear from \fref{fig:quartmot}, in order to stabilize a
structure with a certain spacing between the particles one needs
interactions that can reach at least to such a distance. In
particular for cold atoms, since the interactions are mostly local,
one can expect a Mott insulator to be possible for one (or any
integer) number of fermions per site. Other insulating phases (1
boson each two sites etc.) would need longer range interactions.

To study the Mott transition, we thus consider the application of a
periodic potential of period wavevector $Q=2\pi/a$. This can be
realized by taking
\begin{equation} \label{eq:perbos}
 V_L(x) = \sum_n V^0_n \cos(Q n x)
\end{equation}
In fact in cold atomic gases it is easy to realize systems with only
one harmonic as was shown in (\ref{eq:boslat}). In that case $n=1$
is the only existing Fourier component. In the lattice potential is
very large, then as we already discussed the kinetic energy gets
very small and one has a rather trivial insulating case. In the
Bose-Hubbard language this corresponds to the limit $U \gg t$. The
particles are nearly ``classically'' localized. The case where
either the lattice or the interactions are small is much more
subtle.

\subsection{Bosonization solution}

Using (\ref{eq:perbos}) and the expression (\ref{eq:locintbos}) for
the density, we see that terms such as
\begin{equation} \label{eq:oscbase}
 \int dx e^{i (Q n  - 2 p \pi \rho_0)x} e^{-i2p\phi(x)}
\end{equation}
appear. Because the field $\phi(x)$ is a smooth field varying slowly
at the scale of the interparticle distance, if oscillating terms
remain in the integral they will average out leading to a negligible
contribution. The corresponding operator would then disappear from
the Hamiltonian. In order for such terms to be relevant, one needs
to have no oscillating terms in (\ref{eq:oscbase}). This occurs if
\begin{equation} \label{eq:comcond}
 n Q = 2 \pi \rho_0 p
\end{equation}
If we use $\rho_0 = 1/d$ where $d$ is the distance between particles
one has
\begin{equation} \label{eq:umkcond}
 n d = p a
\end{equation}
The corresponding term contributing to the Hamiltonian is
\begin{equation} \label{eq:umkcond1}
 H_L \propto V^0_n \int dx\; \cos(2 p \phi(x))
\end{equation}
The periodic potential has thus changed for commensurate fillings
the simple quadratic hamiltonian (\ref{eq:luthamphen}) of the
Luttinger liquid into a sine-Gordon Hamiltonian
(\ref{eq:luthamphen}) plus (\ref{eq:umkcond1}). This sine-Gordon
Hamiltonian describes in fact in one dimension the physics of any
Mott transition \cite{giamarchi_book_1d}.

Although the term (\ref{eq:umkcond1}) has been derived here for a
weak potential, it appears also in the opposite limit of a strong
barrier if the filling is commensurate, showing that the two limits
are in fact smoothly connected. Indeed if one starts from the
Bose-Hubbard model, the lattice potential $V_L$ is not present
anymore, but the position of the particles is quantized $x_j = j a$
where $j$ is an integer. It means that when one writes the
interaction term one should pay special attention to this when going
to the continuum limit. The fields $\phi$ are smooth so for them one
has $\phi(x_j) \to \phi(x)$ and one can take for them the continuum
limit. This is not the case for the oscillating factors in
(\ref{eq:locintbos}). Such terms are of the form $e^{2p\pi\rho_0
x_j}$. Since $\rho_0 = 1/a$ they oscillate fast and replacing $x_j
\to $ is impossible in such terms. If one was simply doing it, the
fact that the oscillating factors should vanish in order to avoid
the integral over $x$ to be killed would impose for the interaction
term
\begin{equation}
 U \sum_j n_j (n_j-1) \to U \rho_0^2 a \int dx \sum_{p,p'} e^{i(p+p')2\pi\rho_0
 x}e^{-i2(p+p')\phi(x)}
\end{equation}
to choose opposite $p=-p'$ in (\ref{eq:locintbos}) for each of the
densities. This is the normal interaction, that conserves the total
momentum of the particles. However due to the discreteness of $x_j$
other terms are possible. Let us choose for one density $p=0$ ($n_j
\to \rho_0$) and for the other one keep the term with $p$. Then the
interaction term becomes
\begin{equation} \label{eq:limium}
 U a^2 \rho_0^2 \sum_j e^{2p\pi\rho_0 x_j} e^{-i2p\phi(x_j)} \to
 U a^2 \rho_0^2 \sum_j e^{2p\pi\rho_0 x_j} e^{-i2p\phi(x)}
\end{equation}
Now normally such terms would be killed by the oscillating factor,
but if $2 p \pi \rho_0 a = 2 \pi n$ then the exponential term is
always one, and the corresponding interaction remains in the
continuum limit
\begin{equation}
 U \rho_0^2 a \int dx e^{-i2p\phi(x)}
\end{equation}
which is exactly the same condition and operator than the ones
leading to (\ref{eq:umkcond1}). On a physical basis, these
interactions, known as the umklapp process
\cite{dzyaloshinskii_umklapp}, do not conserve the momentum. However
on a lattice momentum needs only to be conserved modulo one vector
of the reciprocal lattice, the extra momentum being transferred as a
whole to the periodic structure. Note that the main difference
between the weak and strong lattices is the strength of this umklapp
process. For the weak lattice (\ref{eq:umkcond1}) the strength of
the umklapp is simply the amplitude of the periodic potential. For
very large lattice, the umklapp strength becomes now proportional to
the interaction $U$. Of course such a representation works if the
interaction remains reasonably weak compared to the kinetic energy
$t$, otherwise the amplitudes of the operators cannot be determined
directly as discussed above.

Doping causes a slight deviation from the condition
(\ref{eq:umkcond}). This can be seen in two ways. The simplest is to
use the fact that the density is slightly different than the
commensurate density that leads to the relation (\ref{eq:umkcond1}).
Part of the oscillating term remain, but if the deviation is quite
small these oscillations will only be important at very large
lengthscales. One should thus keep the corresponding term. In that
case the umklapp term (\ref{eq:limium}) becomes
\begin{equation} \label{eq:complet}
 H_u = g_u \int dx\; \cos(2 p\phi(x) - \delta x)
\end{equation}
where $p$ is the order of the commensurability and $\delta$ is the
doping, i.e. the deviation of the density from the commensurate
value. Another way to recover this result is to start from the
commensurate case and apply a chemical potential. Using the boson
representation (\ref{eq:locintbos}) the chemical potential term
becomes
\begin{equation} \label{eq:chemcont}
 \mu_0 \int dx \frac1\pi \nabla\phi(x)
\end{equation}
The chemical potential can be absorbed by a redefinition of the
field $\phi$. Introducing
\begin{equation}
\begin{split}
 \tilde\phi(x) &= \phi(x) + \frac{K}{u}\mu_0 x \\
 \tilde\theta(x) &= \theta(x)
\end{split}
\end{equation}
the Hamiltonian is now quadratic again in $\tilde\phi$ while the
commensurate umklapp (\ref{eq:umkcond}) is now changed into
(\ref{eq:complet}). We see again that an incommensurate filling is
washing out the cosine, therefore leading back to a Luttinger liquid
state. However if the deviations from commensurability are small the
doping is only acting for lengthscales larger than $1/\delta$ that
can be quite large compared to the lattice spacing. This leads to an
interesting physics that I examine below. For more details on the
Mott transition and the difference between working with a fixed
density and a fixed chemical potential see \cite{giamarchi_book_1d}.
The Hamiltonian (\ref{eq:complet}) thus provides a complete
description of the Mott transition and the Mott insulating state in
one dimension. To change the physical properties of a commensurate
system one has thus two control parameters. One can vary the
strength of the interactions while staying at commensurate filling,
or vary the chemical potential (or filling) while keeping the
interactions constant. One can thus expect two different classes of
transition to occur.

\begin{figure}
  \centerline{\includegraphics[width=\normfig]{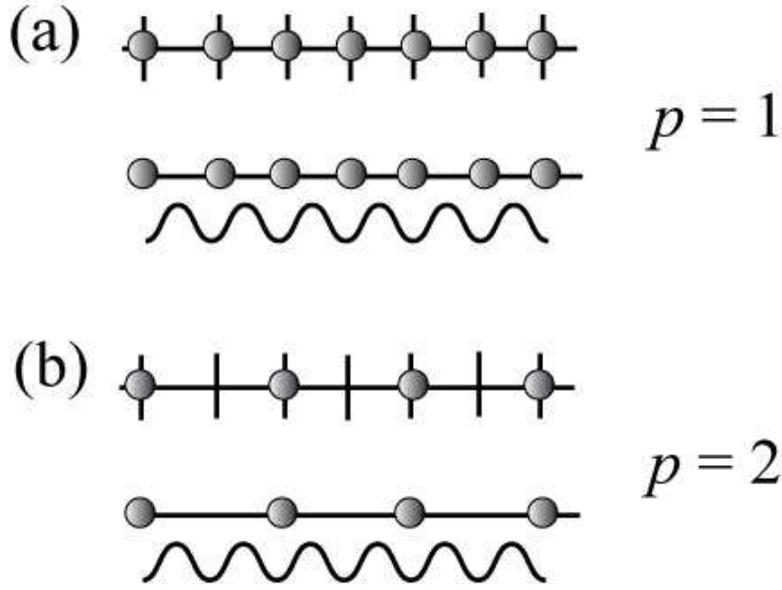}}
 \caption{Commensurabilities for the boson system. (Bottom) Bosons in the continuum. The
 lattice is reintroduced as a periodic potential. (Top) The bosons are defined directly
 on a lattice. The two descriptions lead to the same physics. (a) A commensurability
 of one boson per site. For the periodic potential it means that the density modulation has
 the same period than the external periodic potential. This is the only insulating phase that can
 be stabilized with local interactions. (b) A commensurability of one boson every
 two sites. In that case the period of the density modulation is twice the one of the external potential.}
 \label{fig:combos}
\end{figure}

\begin{figure}
  \centerline{\includegraphics[width=\normfig]{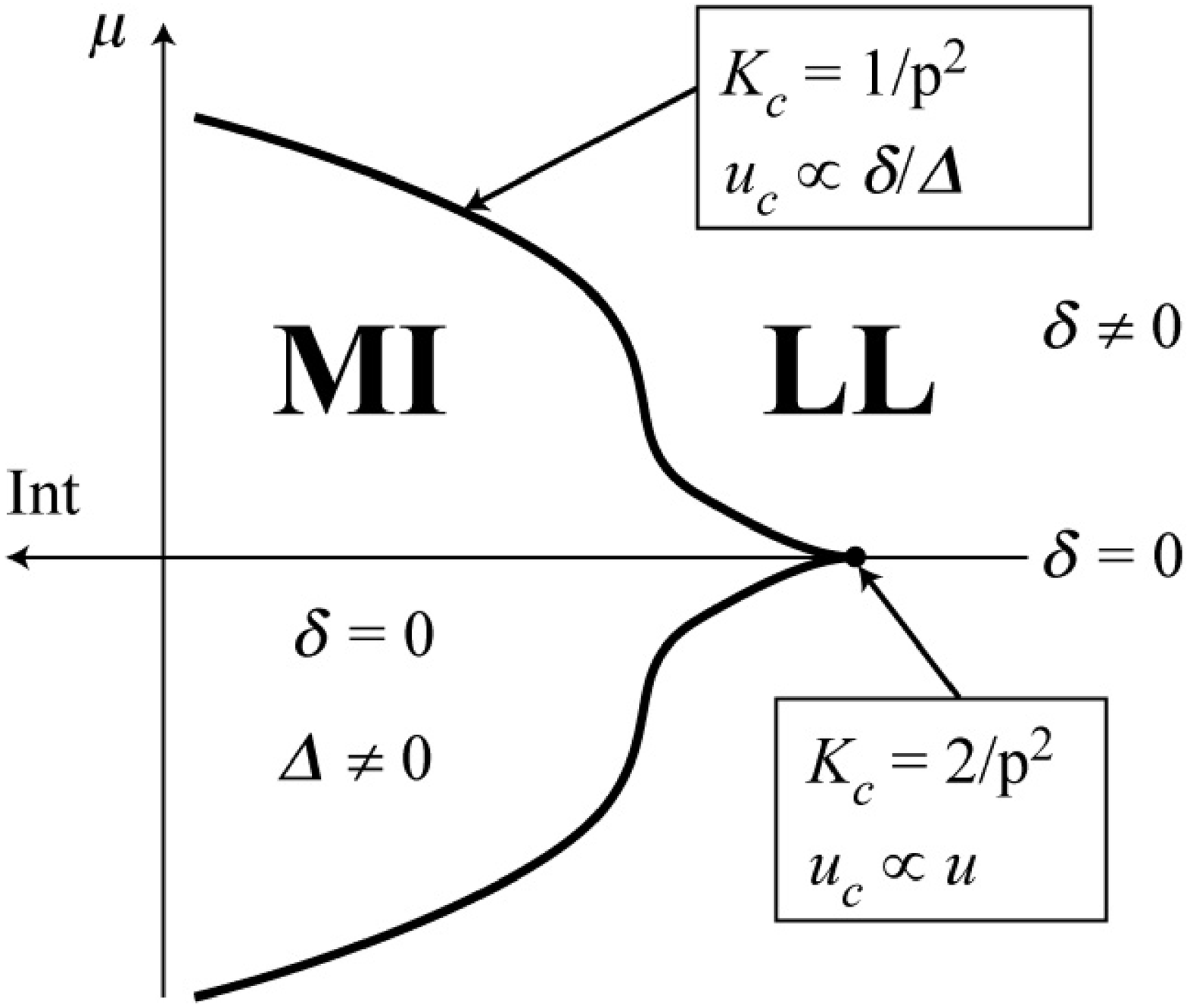}}
 \caption{Phase diagram close to a commensurability of order $p$
 ($p=1$ for one boson per site and $p=2$ for one boson every two sites). Int
 denotes a general (that is, not necessarily local) repulsive interaction. $\mu$ is
 the chemical potential, $\delta$ the doping and $\Delta$ the Mott gap.
 MI and LL are respectively the Mott insulator the Luttinger liquid (massless) phases. The critical
 exponent $K_c$ and velocity $u_c$ at the transition depend on whether it is a Mott-$U$
 or Mott-$\delta$ transition. (After \figcite{giamarchi_mott_shortrev}.)}
 \label{fig:phasediagmott}
\end{figure}

Let us first deal with the transition where the filling is kept
commensurate and interaction strength is varied (Mott-U transition).
In that case $\delta=0$ and (\ref{eq:complet}) is just a sine-Gordon
Hamiltonian. As is well known this Hamiltonian has a quantum phase
transition at $T=0$ as a function of the Luttinger parameter $K$,
and thus as a function of the strength (and range) of the
interactions. This transition is a Berezinskii--Kosterlitz--Thouless
(BKT) transition \cite{kosterlitz_renormalisation_xy}. I will not
explain here how to analyze such a transition, but simply remind how
one can get renormalization equations giving the phase diagram. The
idea is to vary the cutoff $\alpha$ of the theory to eliminate short
distance degrees of freedom, and capture the large distance physics.
Parametrizing the cutoff as $\alpha(l) = \alpha e^{l}$ one can
establish how the parameters in the Hamiltonian must vary when $l$
is varied in order to keep the long distance physics invariant. The
renormalization equations for $K$ and the strength of the umklapp
term (\ref{eq:umkcond1}) are
\begin{equation} \label{eq:rgmott}
\begin{split}
 \frac{dK}{dl} &= -\frac{p^2 K^2}{8\pi\alpha^2 u}  (V^0_n)^2 \\
 \frac{d V^0_n}{dl} &= (2 - p^2 K) V^0_n
\end{split}
\end{equation}
The second equation can be understood by looking at the scaling
dimension of the second order perturbation theory in
(\ref{eq:umkcond1}). Such a term behaves as
\begin{equation}
 (V^0_n)^2 \int dxd\tau \int dx'd\tau' \langle
 T_\tau e^{i2p\phi(x,\tau)}e^{-i2p\phi(x,\tau)} \rangle
\end{equation}
Using the fact that the correlation decay as a power law with an
exponent $2 p^2 K$ the scaling dimension of this integral is
$L^{4-2p^2 K}$. Such dimension leads directly to the second equation
in (\ref{eq:rgmott}). The first equation is more subtle to obtain
\cite{giamarchi_book_1d}. Clearly these equations define two regions
of parameters. If $K$ is large, $V^0_n$ decrease when $l$ increases
which means that the periodic potential is less and less important.
On the contrary, if $K$ is small, $V^0_n$ increases and the cosine
terms is more and more relevant in the Hamiltonian. The critical
value is $K_c = 2/p^2$ where $p$ is the order of the
commensurability. For larger values of $K$ the cosine is irrelevant
and the system is massless. For $K < K_c$ the cosine is relevant and
the system is massive. This opening of a gap corresponds to the Mott
transition and the system becomes an insulator. The larger the
commensurability the smaller $K$ needs to be for the system to
become insulating. From (\ref{eq:umkcond}) we see that for the
bosons $p=1$ corresponds to a commensurability of one (or 2, 3, ...,
which would correspond to higher $n$) boson per site ($n=1$). This
is shown in \fref{fig:combos}.
In that case the critical value is $K_c=2$, which corresponds to
strong but finite repulsion. This means that, contrarily to the
higher dimensional case above a certain threshold of
\emph{interactions} even a \emph{arbitrary weak} lattice will lead
to a Mott insulator. This is a very surprising result, and quite
different from our intuition or the behavior in higher dimension
where one only gets a Mott insulator when the kinetic energy is
small. For one boson each two sites one has $p=2$ as shown in
\fref{fig:combos}. The critical value is $K_c = 1/2$. As discussed
this cannot be reached for a local interactions, but nearest
neighbor repulsion allows to reach this value and to get a Mott
phase. Since for local interactions $1 \leq K < \infty$, one
recovers, directly from the Luttinger theory the argument that one
cannot obtain an ordered phase with a separation of the particles
larger than the range of the interaction. The critical properties of
the transitions are the ones of the BKT transition: $K$ jumps
discontinuously from the universal value $K_c$ at the transition in
the superfluid (non-gapped) regime to zero in the Mott phase (since
there is a gap). Since the velocity is not renormalized it means
using that the compressibility goes to a constant at the transition
and then drops discontinuously to zero inside the Mott phase. A
summary of the critical properties of the Mott transition is given
in \fref{fig:phasediagmott}.
In the Mott phase the single-particle Green's function decays
exponentially since the field $\theta$ is dual to the field $\phi$
which is ordered. The characteristic length of decay is $\xi =
u/\Delta$ where $\Delta$ is the Mott gap. At the transition the
single-particle Green's function decays with a \emph{universal}
exponent $1/(2K_c)$ ($1/4$ for one boson per site, $1$ for one boson
every two sites, etc.). Note that in the LL phase the system is a
perfect conductor (superfluid). A measure is given by the charge
stiffness that is the Drude part of the conductivity $\sigma(\omega)
= \DD \delta(\omega)$. A finite charge stiffness means thus a
perfect conductor for dc transport. The charge stiffness of the LL
is finite $\DD = u_c K_c$. It jumps discontinuously to zero at the
Mott-$U$ transition. In the Mott phase, the system is
incompressible. For more details see \cite{giamarchi_book_1d}.

\begin{figure}
    \centerline{\includegraphics[width=\normfig]{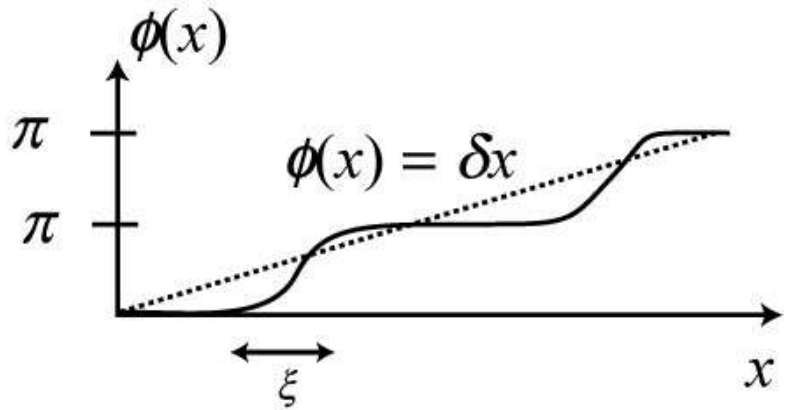}}
\caption{Profile of the field $\phi(x)$ in presence of a
commensurate potential and a finite doping $\delta$. In the absence
of commensurate potential the doping would impose a slope $\phi(x) =
\delta x$ (dashed line). In the presence of the commensurate
potential $\cos(2\phi(x))$ it is more favorable energetically to
maintain commensurability as much as possible and to proceed from
one of the minima of the cosine to the next by making a soliton
(full line). The size of such solitons is $\xi \propto u/\Delta$
where $\Delta$ is the Mott gap. These soliton behave for very small
doping as spinless fermions.} \label{fig:thiring}
\end{figure}

Let me briefly comment on the physics of the doped system
(Mott-$\delta$ transition) \cite{giamarchi_umklapp_1d}. As can be
seen from (\ref{eq:complet}) the doping destroys the cosine and thus
the Mott phase. It is clear that the oscillating term will kill the
cosine at a lengthscale of order $1/\delta$. One has the competition
between two terms: the cosine that would like to keep $\phi$ as
constant as possible and the doping (or the chemical potential) that
would like to tilt $\phi$ so that $\phi = K \mu_0 x/u$ as can be
seen from (\ref{eq:chemcont}). The way this competition takes place
is not to give an homogeneous slope to $\phi$, but to keep $\phi$
commensurate (i.e. locked into one of the minima of the cosine) over
a region of order $1/\delta$ and then create a soliton connecting
two adjacent minima of the cosine. This is shown in
\fref{fig:thiring}. These solitons act in fact like spinless
fermions with some interaction between them. This can be seen by
mapping the sine-Gordon Hamiltonian (\ref{eq:luthamphen}) plus
(\ref{eq:complet}) to a spinless fermion model (known as massive
Thiring model \cite{giamarchi_book_1d}). The remarkable fact is that
\emph{close} to the Mott-$\delta$ transition the solitons become
non-interacting, and one is simply led to a simple semi-conductor
picture of two bands separated by a gap (see \fref{fig:thiring}).
The Mott-$\delta$ transition is thus of the
commensurate-incommensurate type
\cite{japaridze_cic_transition,pokrovsky_talapov_prl,schulz_cic2d,haldane_cic_sinegordon}.
This image has to be used with caution since the solitons are only
non-interacting for infinitesimal doping (or for a very special
value of the initial interaction) and has to be supplemented by
other techniques \cite{giamarchi_umklapp_1d}. Nevertheless it
provides a very appealing description of the excitations and a good
guide to understand the phase diagram and transport properties. The
transition by varying the chemical potential occurs when the
chemical potential equals the charge gap. The density in the
incommensurate phase varies as $n \sim (\mu-\mu_c)^{1/2}$ . The
\emph{universal} (independent of the interactions) value of the
exponents $K_c^\delta = 1/p^2$ is half of the one of Mott-U
transition, as shown in \fref{fig:phasediagmott}. Since at the
Mott-$\delta$ transition the chemical potential is at the bottom of
a band the velocity goes to zero with doping. This leads to a
continuous vanishing of the charge stiffness $\DD \sim
\delta/\Delta$ where $\delta$ is the doping and $\Delta$ the Mott
gap and a divergent compressibility. For more details see
\cite{giamarchi_book_1d}.

\subsection{Extensions} \label{sec:quasi}

As we saw in the previous section, the fact that interactions are
able to lead to a Mott insulator phase has several important
consequences. We have now a fairly good understanding of the
properties of this phase in the pure and homogeneous one dimensional
case. There are of course many open questions and active research
subjects connected to this problem. There are too numerous to be all
mentioned here, so I will simply briefly mention two of them.

First, in cold atomic gases, in addition to the optical lattice
there is usually the harmonic confining potential
(\ref{eq:varchem}). As was already discussed, it acts as a chemical
potential. The density is thus non-uniform, and there is thus no
meaning as looking at the system as wholly in a commensurate Mott
state or not. However as we saw, upon small doping, a Mott insulator
prefers to keep the commensurability as much as possible and makes
discomensurations between two commensurate regions. In presence of
the trap one can thus expect a similar behavior, and to have
sequences of incommensurate regions separated by commensurate ones.
How these regions are organized is an interesting question, that has
been intensely studied
\cite{rigol_groundstate_hcbosons,kollath_dmrg_bose_hubbard_trap,wessel_MC_confined_bosons,%
pedri_tonks_harmonic,Gangardt_correlations}. Another important
question connected to this problem is how to probe for the existence
of such an insulating state. As discussed before measuring the
momentum distribution $n(k)$ gives direct information, since it has
a divergence at $q=0$ for a superfluid phase and none in the
insulating one
\cite{greiner_mott_bec,stoferle_tonks_optical,richard_1dbec_momentum}.
However, $n(k)$ is only providing limited information, and would
also be much less informative in the case of fermion where one has
essentially a broadened step at the Fermi energy regardless of
whether the system is superconducting or insulating. It is thus
important to study other probes of the Mott phase such as noise
\cite{altman_noise_correlations,folling_noise_correlations} or
shaking of the lattice
\cite{stoferle_tonks_optical,schori_absorption_optical}.
Understanding the physics of such shaken lattices is an interesting
problem for which I refer the reader to the literature
\cite{batrouni_dynamic_response,reischl_temperature_mott,iucci_absorption,kollath_shaking_dmrg}.

Another interesting class of problems is to determine how the 1D
Mott insulating properties can affect the physics of the system when
there is not a single chain but many chains coupled together. More
generally it is important to determine how the one-dimensional
physics is changed when one goes from a purely one-dimensional
system to a two- or three-dimensional situation. Such a crossover
between the one dimensional properties and the three dimensional
ones is particularly important since many systems are made of
coupled one dimensional chains
\cite{giamarchi_book_1d,giamarchi_review_chemrev}. Cold atomic
systems provide a very controlled way to probe for such a physics,
since it is possible to control the strength of the optical lattice
in each direction.

If the transverse optical lattice is large it can be treated by the
same tight binding approximation than the one leading to
(\ref{eq:bosehub}). The most important term describing the coupling
between the chains is the interchain tunnelling traducing the fact
that single particles are able to hop from one chain to the next
\begin{equation} \label{eq:singlehop}
 H_\perp = -\int dx \sum_{\langle\mu,\nu\rangle} t_{\perp,\mu,\nu}
 [\psi^\dagger_\mu(x)\psi^\phd_\nu(x) + \hc]
\end{equation}
where $\langle\mu,\nu\rangle$ denotes a pair of chains, and
$t_{\perp,\mu,\nu}$ is the hopping integral between these two
chains. These hopping integrals are of course directly determined by
the overlap of the orbitals of the various chains. In addition to
the single particle hopping, there are of course also in principle
direct interactions terms between the chains. Such terms can be
density-density or spin-spin exchange. However they are easy to
treat using mean field approximation. For example a spin-spin term
$S_\mu S_\nu$ can be viewed, in a mean field approximation, as an
effective `classical' field acting on chain $\nu$: $S_\mu S_\nu \to
\langle S_\mu \rangle S_\nu$. Thus, at least for an infinite number
of chains for which one could expect a mean field approach to be
qualitatively correct, the physics of such a term is transparent: it
pushes the system to an ordered state. Note that for cold atoms
since the interactions are short range such terms do not normally
exist and (\ref{eq:singlehop}) is the only term coupling the chains.
For other cases see \cite{giamarchi_book_1d}.

The single particle hopping is more subtle to treat. For fermions no
mean field description is possible since a single fermion operator
has no classical limit. It is thus impossible to approximate
$\psi^\dagger_\mu(x) \psi^\phd_\nu(x)$ as $\langle
\psi^\dagger_\mu(x) \rangle \psi^\phd_\nu(x)$, which makes the
solution of the problem of coupled chains quite complicated
\cite{arrigoni_tperp_resummation_prb,georges_organics_dinfiplusone,%
biermann_dmft1d_hubbard_short,cazalilla_coupled_fermions}. For
bosons one is in a slightly better situation since the single boson
operator has a mean field value. However even in that case there is
a direct competition between this interchain hopping that would like
to stabilize a three dimensional superfluid phase and the 1D Mott
insulating term that would favor an insulating state. As a function
of the strength of the interchain hopping there is thus a
deconfinement transition where the system goes from a 1D insulator
made of essentially uncoupled chains, to a an anisotropic 3D
superfluid. Such a transition has been studied both theoretically
\cite{ho_deconfinement_coldatoms,cazalilla_deconfinement_long} and
experimentally \cite{stoferle_tonks_optical,koehl_SFMI} and i refer
the reader to these references for more details. Quite generally
such a transition is relevant in various other type of systems as
well \cite{giamarchi_book_1d,giamarchi_review_chemrev}.

\section{Disorder effects: Bose glass} \label{sec:dis}

We have examined in the previous section the effects of a periodic
potential on an interacting bosonic system. Another important class
of potential, leading to radically new physics, is the case of a
disordered potential. Here again the bosonization solution is a
powerful tool to tackle this problem.

\subsection{Disorder in quantum systems}

\begin{figure}
 \centerline{\includegraphics[width=\normfig]{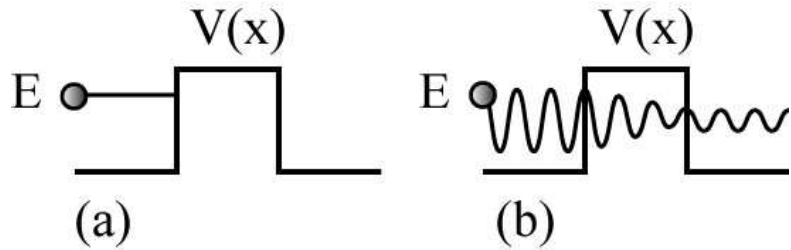}}
\caption{\label{fig:disnaif} (a) A classical particle of energy $E$
smaller than the disorder $V(x)$ at a given point is totally
blocked. (b) A quantum mechanical system of the same energy can pass
through the barrier by tunnel effect. One could therefore naively
think that quantum systems are much less localized by disorder than
their classical equivalents. In fact it is exactly the contrary that
happens.}
\end{figure}

Disorder for quantum problems is a longstanding problem. In
condensed matter, some level of disorder is unavoidable, and it is
thus necessary to deal with it. The naive expectation is to think
that the disorder will have weaker effects for a quantum system than
for a classical one. Indeed, as shown in \fref{fig:disnaif}
one can imagine that waves in a quantum system have more ease to
pass the barriers induced by the disorder since they can use tunnel
effect. It was thus a major surprise when Anderson showed
\cite{anderson_localisation} that it was in fact exactly the
opposite effect that occurred for non-interacting quantum particles.
Indeed because of the constructive interferences of two paths that
are deduced by time inversion there is an additional probability for
a particle to be backscattered by the disorder
\cite{bergmann_weak_localization}. Loosely speaking one should add
the wave functions, and thus squaring them getting a factor of four,
instead of the naive factor of two of two paths that would not
interfere. The main effect is that the wavefunctions of the system,
instead of being plane waves, now decay exponentially in space. This
phenomenon, known as Anderson localization is strongly dependent on
dimension. Simple scaling arguments show that all states should be
localized in one and two dimensions \cite{abrahams_loc}. In three
dimensions, a mobility edge in energy exists below which states are
localized and above which they are extended. An important
characteristic of such states is thus the localization length $\xi$
characterizing the spatial decay of the localized states. This
phenomenon is now well understood for noninteracting particles. For
Fermions, i.e. electrons is condensed matter, interactions do exist.
However because of the Pauli principle, the important electrons, at
the Fermi level have a large kinetic energy $E_F$. If this energy is
large compared to their interaction it is very reasonable to assume
that the noninteracting limit is a good starting point. Indeed, the
corresponding predictions for the localization localization have
been spectacularly confirmed experimentally
\cite{bergmann_weak_localization}.

Treating the combined effects of interactions and disorder is a
particularly challenging problem, even for fermions. Indeed because
of the disorder the motion of the particles becomes much slower than
the one of free particles. From ballistic it becomes diffusive at
best, which means that two particles can spend more time close to
each other. There is thus an extremely strong reinforcement of the
interactions by the disorder
\cite{altshuler_aronov,finkelstein_localization_interactions}. This
leads to singularities and to a physics that is still under debate.
Here again the effect of the dimension is crucial, since the
singularities increase with lowering dimension. One can expect one
dimension, where disorder lead to all state being localized and the
interaction leads to the Luttinger liquid state, to be particularly
special. I will not dwell further on this problem here and refer the
reader to the above literature for further references.

\begin{figure}
  \centerline{\includegraphics[width=\normfig]{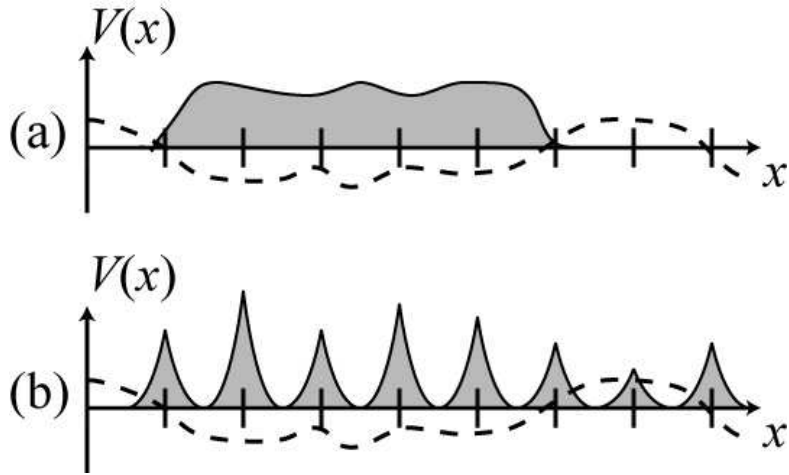}}
 \caption{A cartoon of free versus interacting bosons in the
 presence of disorder (represented by the dashed line). (a) Free boson all condense in a large enough well of the random
 potential. Since the density is infinite, this situation is unstable when interactions are added.
 (b) the repulsion prevents the bosons to condense in the same site. Their behavior is thus
 much closer to fermions, where various minima of the random potential have to be used. Since one
 can still pile up many bosons in the same minima, the deep minima of the random potential are
 well smoothed by the bosons.}
 \label{fig:raredis}
\end{figure}

The case of bosons is even more interesting. Indeed in that case the
noninteracting case cannot even be used as a reasonable starting
point. To understand this let us simply look at a disorder that
would take two values $\pm V_0$ on each site. Let us assume that one
can find a region of space of length $L_0$, as shown in
\fref{fig:raredis}. Such a region always exists with a probability
$e^{-L_0/L_c}$ where $L_c$ is the characteristic correlation length
of the disorder.
The lowest energy of one boson confined to this region can be
readily computed. Because the boson is confined to a region of size
$L_0$ its momentum is $p \sim \pi/L_0$ instead of zero, since the
wavefunction has to essentially vanish at the edges of the region.
Thus the total energy is
\begin{equation}
 \Delta E = \frac{1}{2m}\left(\frac{\pi}{L_0}\right)^2 - V_0
\end{equation}
It is thus easy to see that provided that $L_0$ is large enough this
energy is lower than putting the boson in the lowest plane wave
state with $p=0$ where the kinetic energy and average disorder
energy would both be zero. Thus one boson will simply go into this
finite size region. But then for noninteracting bosons they all
condense in the same state. For noninteracting bosons, the
superfluid is thus destroyed for arbitrarily weak disorder, and all
the particles go to a region of finite size, forming a puddle. The
effect of the interactions on such a state is of course crucial,
since a macroscopic number of particles $N$ (proportional to the
total size of the system $L$) condense into a finite size region
$L_0$ the density is infinite. Any infinitesimal repulsion makes
thus this state unstable. For bosons one has thus to include
interactions from the start to get a meaningful answer.

\subsection{Disordered interacting bosons}

Let us now turn to the problem of such an interacting disordered
bosonic gas. In one dimension this problem was solved in
\cite{giamarchi_loc}. Building on this microscopic solution scaling
analysis were developped to investigate this question in higher
dimensions \cite{fisher_boson_loc}. Here also the bosonization
representation is particularly useful to deal with the effects of
disorder on the one-dimensional boson gas. The disorder can be
introduced as a random potential coupled to the density. For
simplicity I stick here to the incommensurate case. The disorder is
\begin{equation}
 H_{\rm dis} = \int dx\; V(x) \rho(x)
\end{equation}
where $V(x)$ is a random variable. One should fix the distribution
for $V(x)$ which of course depends on the problem at hand. However
if the disorder is weak so that the characteristics of the boson
system vary slowly at the lengthscale of variation of the disorder,
central limit theorem shows that one can approximate the
distribution by a gaussian one. Using the representation
(\ref{eq:locintbos}) of the density one has (keeping only the
lowest, that is, most relevant harmonics)
\begin{equation}
 H_{\rm dis} = \int dx\; V(x) [-\frac1\pi\nabla\phi(x) +
 \rho_0(e^{i(2\pi\rho_0 x - 2 \phi(x))} + \hc)]
\end{equation}
This expression shows one remarkable fact. Different Fourier
components of the disorder act quite differently on the density, and
it is important to distinguish these Fourier components. The natural
separation between these different terms is again $Q \sim 2 \pi
\rho_0$, i.e. the average distance between the bosons.

\begin{figure}
 \centerline{\includegraphics[width=\normfig]{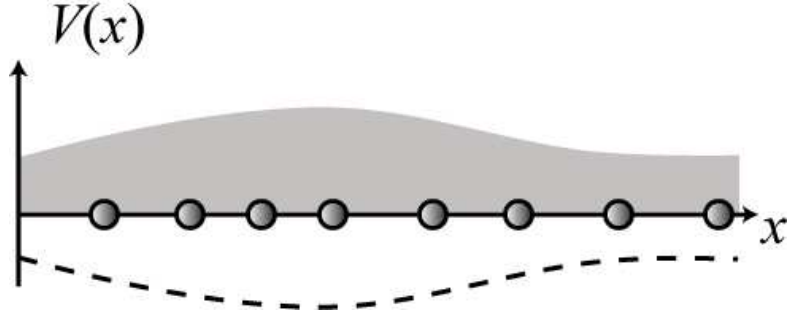}}
\caption{\label{fig:forward} The Fourier component of the random
potential $V(x)$ with wavevector small compared to the inverse
particle distance act as a random chemical potential. This term
increases or decreases smoothly the density. For weak disorder this
term does not lead to any localization of the bosons and does not
affect the currents or the superfluid correlations. For the TG
limit, it would be the equivalent of the forward scattering for the
Fermions (see text). The density is indicated by the gray area,
while a schematic position of the bosons is given.}
\end{figure}

The first term is
\begin{equation}
 H_f = -\int dx\; V(x) \frac1\pi\nabla\phi(x)
\end{equation}
since the field $\phi$ is smooth at the length scale of the distance
between particles, this term couples essentially to the smooth
variations of the disorder $V(x)$ varying at a lengthscale much
larger than the distance between particles. Note the analogy with
the chemical potential term (\ref{eq:chemcont}). This term is
analogous so a slowly varying chemical potential. It is easy to see
that this term can be again trivially absorbed
\cite{giamarchi_loc,giamarchi_book_1d} in a redefinition of the
field $\phi$ by
\begin{equation} \label{eq:transf}
 \tilde\phi(x) = \phi(x) + \frac{K}u \int_0^x dy V(y)
\end{equation}
This means that the smeared density $\rho(x) =
-\frac1\pi\nabla\phi(x)$ follows the variation of the potential as
shown in \fref{fig:forward}
Note that the coefficient that relates the change of density to the
change of potential is of course the compressibility of the bosons,
which is now finite due to the interactions. This term is thus a
very classical effect where the bosons go in puddles in the holes of
the random potential.

The oscillations at $q \sim 2\pi\rho_0$ of the density are deeply
affected. In the pure system these correlations were decaying as a
powerlaw. Now they behave as
\begin{equation} \label{eq:forone}
 \langle e^{i2\phi(x,\tau)} e^{-i2\phi(y,\tau')} \rangle \to
 e^{i\frac{2K}u\int_y^x dz V(z)} \langle e^{i2\phi(x,\tau)} e^{-i2\phi(y,\tau')} \rangle_{\rm pure}
\end{equation}
If one takes a gaussian disorder with a distribution
\begin{equation}
p(V(x)) \propto e^{-D_f^{^-1}\int dx V(x)^2}
\end{equation}
which leads to averages such as $\overline{V(x)V(x')} = D_f
\delta(x-x')$, then the average of the expression \ref{eq:forone})
gives
\begin{equation}
 e^{-\frac{D K^2}{u^2} |x-y|} \langle e^{i2\phi(x,\tau)} e^{-i2\phi(y,\tau')} \rangle_{\rm pure}
\end{equation}
leading to an exponential decay of the correlations of the density
waves. This is due to the fact that this disorder introduces a
random phase in the position of the oscillations of the density.

Paradoxically such a term does \emph{not} lead to any localization
of the bosons. Indeed if one computes the current of bosons, it is
given by $J =
\partial_\tau \phi$. The transformation (\ref{eq:transf}) thus
leaves the current invariant and identical to the ones of pure
bosons. From the point of view of the transport the system remains a
superfluid. Note also that the field $\theta$ is unchanged by the
transformation (\ref{eq:transf}), which means that the superfluid
correlations are identical to the ones of the pure system. A
particularly transparent interpretation of this term can be inferred
by looking at the TG limit, or simply at the comparison between the
bosonized expressions for the bosons and the spinless fermions
systems. As already noted in the Tonks regime $2\pi\rho_0 \to 2
k_F$, where $k_F$ is the Fermi wavevector. Such a disorder thus
corresponds to \emph{forward scattering} where a fermion around $\pm
k_F$ remains around the same point of the Fermi surface. It is now
obvious that such a forward scattering cannot essentially change the
current and cannot lead to localization.

\begin{figure}
 \centerline{\includegraphics[width=\normfig]{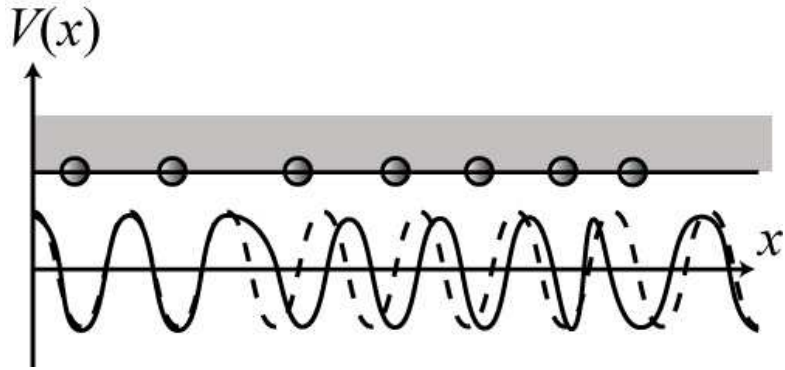}}
\caption{\label{fig:backsimp} The Fourier components of the disorder
$V(x)$ with wavevector close to $Q=2\pi\rho_0$ is the one
responsible for the localization and the formation of the Bose
glass. The full line shows the disorder, while the dashed line would
be a periodic potential of wavevector $Q$. The disorder matches the
periodicity of the bosons, thus acting in a similar way than a Mott
potential, but different portions of the system are shifted compared
to the perfect periodic position. Thus although there is pinning of
the bosons, the density arrangement is not perfect. Note that for
this disorder the smeared density at wavelengthes much smaller than
$Q$ is essentially constant as indicated by the gray box. This
mechanism of localization is thus quite different than the simple
formation of puddles of bosons.}
\end{figure}

Much more interesting effects arise from the other term, namely
\begin{equation} \label{eq:disback}
 H_{\rm dis} = \int dx\; V(x)
 \rho_0(e^{i(2\pi\rho_0 x - 2 \phi(x))} + \hc)]
\end{equation}
Because $\phi$ is a smooth field it is easy to see that this term
corresponds now to coupling of Fourier components of the disorder
with components around $q \sim \pm 2\pi\rho_0$. One can thus rewrite
\begin{eqnarray}
 V(x) &=& \sum_k V_k e^{i k x} \nonumber \\
      &=& \sum_{|q| \ll Q} V_{Q + q} e^{i (q +Q) x} + \hc \nonumber \\
      &=& e^{i Q x} \xi(x) + \hc
\end{eqnarray}
where $Q = 2\pi\rho_0$ and the above equation defines $\xi(x)$.
Contrarily to $V(x)$ $\xi(x)$ is a smooth field with averages over
disorder of the form
\begin{equation}
\begin{split}
 \overline{\xi(x)\xi^*(y)} &= \dback \delta(x-y) \\
 \overline{\xi(x)\xi(y)} &= 0
\end{split}
\end{equation}
If $\xi(x)$ was simply constant, (\ref{eq:disback}) would correspond
to a commensurate periodic potential, and one would be back to the
case of the Mott transition explained in the previous section. The
fact that $V(x)$ is random makes the phase of the periodic
modulation vary from different positions. We thus see that for the
quantum system of bosons, this component of the disorder acts a
little bit in a similar way than a periodic potential, trying to pin
the charge density wave of bosons. However, because of these phases
fluctuations the pinning is not perfect and varies from place to
place, leading to a distorted charge modulation. This is shown in
\fref{fig:backsimp}.
One can also get a simple interpretation for this term by going to
the Tonks limit. Indeed in that case this term represents a
scattering by the disorder with a momentum close to $\pm 2 k_F$. It
is thus a backscattering term, where a right moving fermion is
transformed into a left moving one and vice versa. It is thus clear
that such a term affects the current. Exact solutions for
noninteracting fermions indeed shows that this term is the one
responsible for Anderson localization.

In order to solve for the generic boson system, we can, as for the
Mott transition, write the renormalization equations for the
disorder and the interactions. The procedure to obtain them is
detailed in \cite{giamarchi_loc,giamarchi_book_1d}. One finds.
\begin{equation}
\begin{split}
 \frac{dK}{dl} &= -\frac{K^2}2  \dbacktilde \\
 \frac{d D}{dl} &= (3-2K) \dbacktilde
\end{split}
\end{equation}
where $\dbacktilde = \dback/(\pi^2 u^2 \rho_0)$ and $\dback$ is the
backward scattering.

The phase diagram can be extracted from these equations, exactly in
the same spirit than what was done for the Mott transition in the
previous section. The disorder is irrelevant for $K>3/2$, that is,
weakly repulsive bosons. One finds a localized phase for $K<3/2$,
that is, if the repulsion between the bosons is strong enough. On
the separatrix between the two phases the parameter $K$ takes the
universal value $K^*=3/2$. Thus, the correlation functions decay
with \emph{universal} exponents. For example, the single-particle
correlation function decays with an exponent $1/3$. This calculation
thus point out the existence for the bosons of a localized phase.
This phase, nicknamed Bose glass, whose existence can be established
microscopically in one dimension \cite{giamarchi_loc} has been
generalizable to higher dimensions as well \cite{fisher_boson_loc}.
In one dimension, one can compute the critical properties of the
transition between the superfluid and the Bose glass. I refer the
reader to \cite{giamarchi_loc,giamarchi_book_1d} for more details on
that point. In particular the superfluid stiffness $\DD$ jumps
discontinuously to zero in the Bose glass phase and $\DD/u = K$
takes the universal value $3/2$ at the transition. At the transition
the disorder is marginal. Because of the dual nature of the phases
$\phi$ and $\theta$ the fact that the phase $\phi$ is now pinned
means that the superfluid correlations decay exponentially, with a
characteristic length that is the localization length. One thus
expects a lorentzian shape for the $n(k)$ instead of the divergent
powerlaw behavior of a Luttinger liquid. The correlation length
diverges at the transition to the superfluid phase. Other methods
can be used to extract information on the localized phase
\cite{giamarchi_quantum_pinning}.

This transition from the superfluid to the Bose glass is a direct
consequence of the interaction effects between the bosons. In
particular the fact that one has a strongly correlated system is
hidden in the conjugation relation between the phase $\phi$ and
$\theta$ which forces the density fluctuations to be directly
related to the superfluid ones. In higher dimensions, although the
excitations of the superfluid phase $\theta$ can be described by
sound waves, this would not imply much for the fluctuations of the
density. In the Bose glass phase, the localization is very similar
to the one for spinless fermions. The bosonic nature of the
particles is not so important any more. To summarize, in order to be
able to observe the localization for quantum interacting bosons, it
is important to fulfill the following conditions
\begin{enumerate}
\item Have a disorder with sizeable Fourier component close to the
interboson periodicity. Having a too smooth potential is of little
help, since it can lead to some puddle separation if the disorder is
strong but this is a very ``classical localization''. The Bose glass
phase can also occur for weak disorder.

\item Have repulsive enough interactions between bosons. If $K <
3/2$ even an infinitesimal disorder is able to localize. Of course
one wants the localization length to be smaller than the size of the
system, to observe the localization. The larger the disorder, the
larger of course the value of $K$ at which the system localizes. One
does not want to make the disorder too strong though (not stronger
than the chemical potential) otherwise one is back to the puddle
localization mentioned above.
\end{enumerate}
How to reach such limits in a realistic cold atomic system is of
course a very challenging question. Current system seem not one
dimensional enough and/or with too smooth disorder to be in this
quantum limit \cite{clement_disorder_quasi_1D}. One is close however
and there is thus little doubts that such a state will be reached in
a near future.

There are many directions in which these questions of disorder
acting on bosonic systems can be further studied. First for the
disordered problem numerical studies have confirmed the analytical
predictions and allowed to further study the phase diagram directly
in terms of the microscopic parameters
\cite{scalettar_bosons,krauth_bosons_disorder,rapsch_bosons_disordered_numerics}.
It is clear that similar studies taking into account the
peculiarities of the system (trap etc.) would be very interesting in
the context of cold atomic gases.

Second, in the previous section we saw the effect of a periodic
potential. We saw that it is very efficient into opening a gap and
leading to a Mott insulating phase, but only if the filling is
commensurate with the periodicity. On the other hand, a random
potential is slightly efficient in giving an insulating phase, but
can act regardless of the density of bosons. A particularly
interesting intermediate case is the case of quasi-periodic
potentials. These potentials lead to a new universality class for
the superfluid-insulator transition
\cite{vidal_quasi_interactions_short,hida_precious}. Similarly one
expects very interesting effects when combining disorder and
commensurability
\cite{fisher_boson_loc,scalettar_bosons,giamarchi_mottglass_long} or
going for more than one bosonic mode inside the tube. I refer for
example the reader to the literature for other examples of
interesting problems such as going to systems of coupled chains
\cite{orignac_2chain_bosonic,donohue_deconfinement_bosons}. It will
be very interesting to see if some of these effects can be directly
tested in a cold atomic gas context.

\section{Conclusions and perspectives} \label{sec:conc}

I have shown in these brief notes some of the properties of
interacting particles in one dimension. I have focussed principally
on interacting one dimensional bosons. Many more examples both on
bosons and on other systems can be found in
\cite{giamarchi_book_1d}. Among the many efficient methods both
analytical and numerical to tackle one dimensional systems, I have
chosen to present here a short account the bosonization method. It
is one of the most versatile and physically transparent method. In
addition to providing direct insight on the low energy properties of
the system, it can also complement very well other angles of
approach such as numerical ones. Here again the avid reader will
find other methods explained in \cite{giamarchi_book_1d}. I have
shown application of this method to two problems of importance in
the rapidly growing field of cold atoms in optical lattices: the
Mott transition induced by the presence of a periodic potential on
interacting bosons, and the localization of interacting bosons in
the presence of a random potential.

Despite an history of more than 40 years the one dimensional world
thus continues to offer fascinating challenges. In that respect cold
atomic gases have opened a cornucopia of possibilities to test for
this fascinating physics. This is due both to the level of control
offered by such system but also by their ability to deal with
bosons, fermions or mixtures of them at will. It is clear that they
have raised many more questions than the theorist had answers ready
for, hence offering new playgrounds and challenges. Under such an
experimental pressure, there is thus little doubts that one can
expect spectacular progress in the years to come.


\begin{theacknowledgments}
Although these notes discuss about one dimensional systems, most of
the material is centered around cold atomic physics. I thus would
like to specially thank M. A. Cazalilla, A. F. Ho, A. Iucci, C.
Kollath, with whom my own research in the field of cold atomic gases
has been conducted, for the many enjoyable discussions and ongoing
collaborations on this subject. This work has been supported in part
by the Swiss National Fund for research under MANEP and Division II.
\end{theacknowledgments}


\bibliographystyle{aipproc}   


\IfFileExists{\jobname.bbl}{}
 {\typeout{}
  \typeout{******************************************}
  \typeout{** Please run "bibtex \jobname" to optain}
  \typeout{** the bibliography and then re-run LaTeX}
  \typeout{** twice to fix the references!}
  \typeout{******************************************}
  \typeout{}
 }

\end{document}